\documentclass[prd,showpacs,amsmath,showkeys,twocolumn,floatfix]{revtex4} 
\usepackage{epsfig,dcolumn}
\usepackage{graphicx}
\usepackage[usenames]{color}
\usepackage{graphicx}
\usepackage{bm}

\def\p{{\bf p}}

\def\lsim{\mathrel{\rlap{\lower4pt\hbox{\hskip1pt$\sim$}}
    \raise1pt\hbox{$<$}}}
\def\gsim{\mathrel{\rlap{\lower4pt\hbox{\hskip1pt$\sim$}}
    \raise1pt\hbox{$>$}}}
\begin{document}

\title{The role of $P$-wave inelasticity in $J/\psi \to \pi^+\pi^-\pi^0$ }

\author{  Peng Guo$^1$,  Ryan Mitchell$^1$  and Adam P. Szczepaniak$^{1,2}$}
\affiliation{ $^1$ Physics Department, Indiana University, Bloomington, IN 47405, USA. \\
$^2$Center For Expiration of Energy and Matter, Indiana University, Bloomington, IN 47408, USA. 
}

\begin{abstract}

We discuss the importance of inelasticity in the $P$-wave  $\pi\pi$ amplitude on the Dalitz distribution of $3\pi$ events in $J/\psi$ decay. The inelasticity, which becomes sizable for $\pi\pi$ masses above $1.4\mbox{ GeV}$, is attributed to $K{\bar K} \to \pi\pi$ re-scattering. We construct an analytical  model for the two-channel  scattering  amplitude and use it to solve the dispersion relation for the isobar  amplitudes that parametrize  the $J/\psi$ decay.  We present comparisons between theoretical predictions for the Dalitz distribution of $3\pi$ events with available experimental data. 
 \end{abstract} 

\pacs{13.25.Gv, 11.55.Fv, 11.80.Et, 11.80.Gw}

\maketitle

\section{Introduction}


One of the most outstanding difficulties of experimental light quark spectroscopy -- like in studies of charmonium decays to light quark mesons at BES~III~\cite{Asner:2008nq} or future studies of photoproduction at GlueX -- is in the disentanglement of overlapping and interfering meson states, which often have widths of several hundreds of $\mbox{MeV}$.  This requires amplitude analyses, where experimental distributions are described by a seies of theoretical amplitudes ( decay amplitudes ) with each amplitude generally multiplied by a freely fit parameter ( production amplitudes).  In the past, decay amplitudes were generally written using the isobar model, {\it i.e.} assuming a multi-particle decay proceeded through a series of two-body resonance decays with the resonance decays usually parametrized as Breit-Wigner amplitudes.  This model, however, is known to violate unitarity.  With high-statistics data samples now available at BES~III and later in GlueX, as well as other current and future experiments, more careful attention must now be paid to the theoretical descriptions of the decay amplitudes, and phenomena such as final-state re-scattering and inelasticity must be considered.

The decay $J/\psi\to \pi^{+}\pi^{-}\pi^{0}$, which is observed to proceed dominantly through $\rho\pi$, provides a simple context in which re-scattering effects can be studied.  Here the $\pi\pi$ system is limited to either $J^{PC}=1^{--}$ ($P$-wave) or $3^{--}$ ($F$-wave).  Neglecting the small $3^{--}$ component, this reaction thus provides clean access to $P$-wave $\pi\pi$ scattering.  The decay $J/\psi\to \pi^{+}\pi^{-}\pi^{0}$ has previously been studied experimentally by BES~II~\cite{Bai:2004jn} and BaBar~\cite{Aubert:2004kj}, but limited statistics prevented any detailed analysis of the $3\pi$ substructure.  BES~III will soon have a set of $J/\psi$ decays many times larger than what is now available, and this data set could be used to greatly improve many of the theoretical uncertainties associated with re-scattering effects.

In this work, we present a coupled channel analysis of $J/\psi \to \pi^+\pi^-\pi^0$ decays in which we consider both $\pi\pi$ and $K\bar K$ isospin-1 intermediate states. In particular, we take advantage of unitarity  constraints to reconstruct the amplitudes based on their analytical properties. Unitarity relates the discontinuity of the isobar amplitude to the scattering amplitude and we use the available data on  $P$-wave $\pi\pi$ scattering to construct analytical $\pi\pi$ and $K{\bar K}$ scattering amplitudes.  We show that available data on the $3\pi$ decay of the $J/\psi$ is  inconsistent  with the single channel parametrization. The effect of the intermediate $K{\bar K}$ pairs is to enhance the contribution from the tail of the $\rho(770)$ while reducing contributions from higher-mass $\rho$ excitations. 

This paper is organized as follows. In the following section, we discuss the analytical properties of the production and scattering amplitudes. We also construct an analytical  model for two-channel $\pi\pi$ and $K\bar K$ scattering and finally compare theoretical predictions with the experimental data.  A summary is  given in Section~\ref{conclusion}.

\section{ $P$-wave $\pi\pi$ effects in  $J/\psi \rightarrow  \pi^{+} \pi^{-} \pi^{0}$ decay } 
\label{sec:P} 
For each helicity state, $\lambda$, of the $J/\psi$, the amplitude to decay to three pions is a function of three angles and two invariant masses. In the rest frame of the $J/\psi$, the angles may be chosen to specify the orientation of the plane formed by the momenta of the three produced pions with respect to the direction of polarization of the $J/\psi$. The invariant masses correspond then to the Dalitz variables describing the  $3\pi$ system. Denoting the four-momenta by $p_{\pm,0}$, $P$ for $\pi^\pm$, $\pi^0$  and $J/\psi$, respectively, the general expression for the amplitude is given by
\begin{equation} 
\langle \pi^0\pi^+\pi^-,out|J/\psi(\lambda),in\rangle =  (2\pi)^4   \delta^4(\sum_{i=0,\pm} p_i - P) 
i T_\lambda,   \label{J}
\end{equation} 
with, in the rest frame  of the $J/\psi$, 
\begin{equation} 
T_\lambda = -i\bm{\epsilon}(\lambda) \cdot (\hat \p_+ \times \hat \p_-) F(s_{+-},s_{0+},s_{-0}). \label{Dalitz}
\end{equation} 
Here $\epsilon$ is the polarization vector of the $J/\psi$,  the Dalitz invariants are defined by $s_{ij} = (p_i + p_j)^2$ for $i,j = \pm,0$ and satisfy $s_{+-} + s_{0+} + s_{-0} = M^2  + 3m_\pi^2$, $\hat \p_i = \p_i/|\p_i|$,  and the scalar form factor $F$ describes the dynamics of the decay. It is $|F|^2$ that  determines the distribution of events in the  Dalitz plot, {\it i.e.} $|F|^2 = const.$ yields a flat distribution.  Since $\sum_i \p_i = 0$ in the $J/\psi$ rest frame, any two pion momenta can be used instead of  $\p_+$ and $\p_-$ in Eq.(\ref{Dalitz})  to specify the orientation of the  decay plane. 

The isobar model makes a specific assumption about $T$, {\it i.e.}  the decay is assumed to proceed   via a quasi two-body process in which a pair of pions in a low partial wave  and a spectator are formed without any  further interactions. The isobar model violates unitarity, which forces interactions between pions from the quasi two-body state and the spectator to be included. If the quasi two-body state, however, is dominated by a low-mass, narrow  resonance, then the overlap between the resonance and the spectator pion  wave functions is expected to be small. Indeed, in the case of the $\pi\pi N$ final state at a  total center of mass energy below $2\mbox{ GeV}$~\cite{Aitchison:1979ja,Aitchison:1979gq} (one of the very few phenomenological analyses of re-scattering effects in three-particle systems that we are aware of), the re-scattering corrections were found to not exceed $20\%$~\cite{Aitchison:1979fj}. In the case of the $J/\psi$ with even higher center of mass energy and with a pronounced $\rho$ resonance in $\pi\pi$,   we expect these effects to be even smaller.  Nevertheless, it will  be important to quantify the size of such re-scattering effects in three-body  $J/\psi$ decays, in particular in view of the very high statistics data currently being collected at BES~III.

\begin{figure}[hh]
\begin{center}
\includegraphics[width=3.5 in,angle=0]{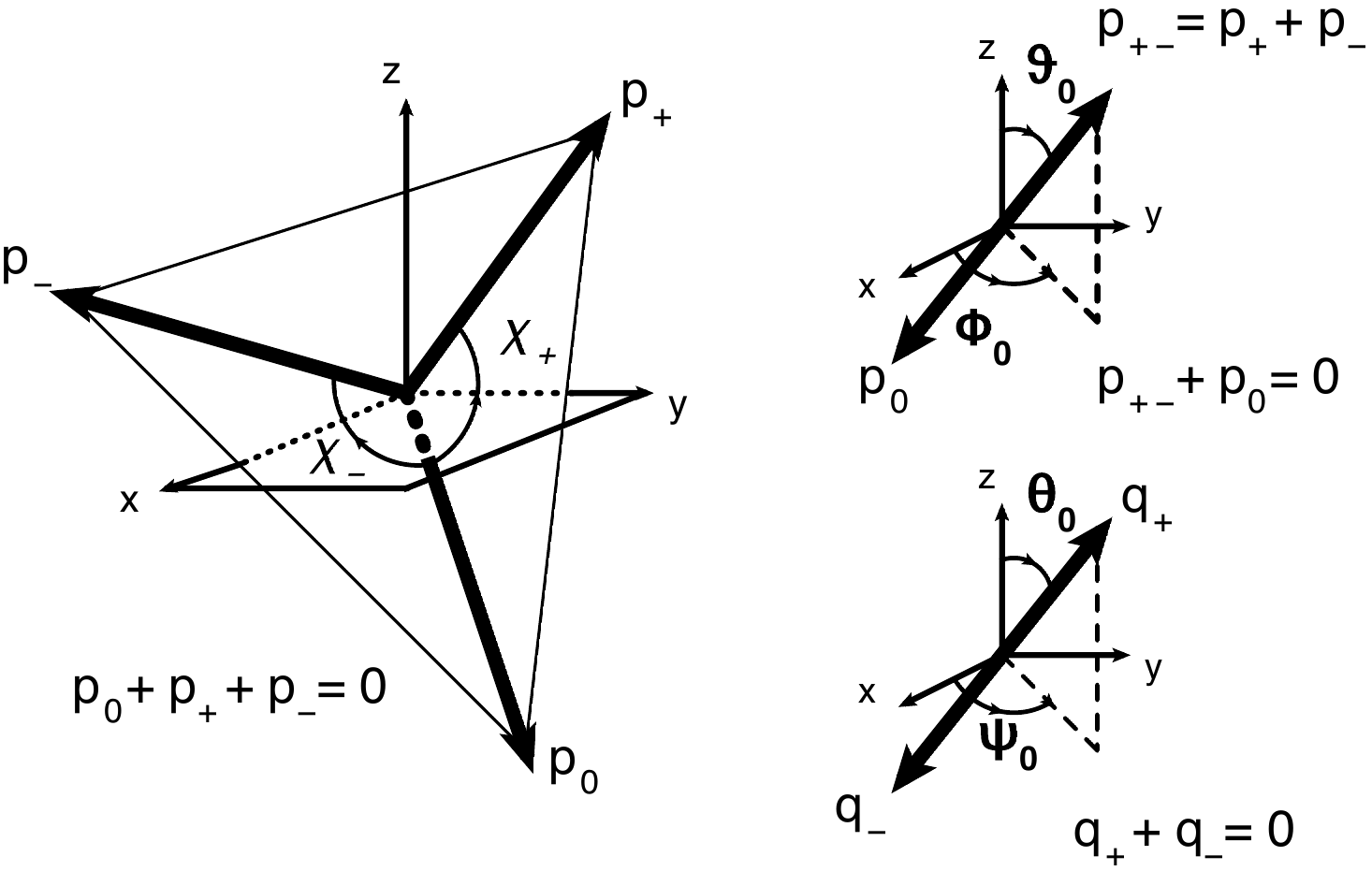}  
\caption{ Definition of the decay angles in the $J/\psi$ rest frame (left and top right) and  the $\pi^+\pi^-$ isobar rest frame (bottom right). 
\label{fig:isobar}}
\end{center}
\end{figure}

The two lowest $\pi\pi$ partial waves allowed in $J/\psi$ decay have $L=1$ (P) and $L=3$ (F).  Little is known about higher partial waves, but the $F$-wave is already very weak with the phase shift staying below $5^0$ for energies up to $1.45\mbox{ GeV}$~\cite{Kaminski:2006yv}. In the following we will thus keep only the $P$-wave in our isobar analysis. Within the isobar model with a single $P$-wave $\pi\pi$ isobar, the amplitude $T$ in Eq.(\ref{J}) is given by 
\begin{equation} 
T_\lambda =  \sum_{i=0,\pm} \sum_{\mu=\pm,0} D^{1*}_{\lambda, \mu}(r_i) d^1_{\mu,0}(\theta_i) F_\mu(s_{jk}) 
\end{equation}  
where the angles are illustrated in Fig.\ref{fig:isobar} and the indices $ijk$ run through cyclic permutations of $0,+,-$~\cite{Brehm:1977yr,brehm:1981}.  
Here $\lambda$ is the spin projection of the $J/\psi$, which, together with the $x$ and $y$ defined with respect to a lab coordinate system, defines the $z$ axis.  
The rotation  $r_k$ is given by three Euler angles, $r_k=r_k(\phi_k,\vartheta_k,\psi_k)$,   which rotates the standard configuration that corresponds to the $(ij)k$ coupling scheme (with the $ij$ forming the $L=1$ isobar and $\pi^k$ being the spectator) to the actual one.   In the  standard configuration $\pi^k$ has momentum along $-z$  and $\pi^i$ and $\pi^j$ have momenta in the $xz$ plane with $\pi^i$ having a positive $x$  component. Finally, $\theta_k$ is the polar angle of the $\pi^i$ in the $\pi^i\pi^j$ rest frame.  
   In other words,  $\phi_k$ and $\vartheta_k$,  are the azimuthal and polar angles, respectively,
   of the total momentum of the $\pi^i\pi^j$  pair in the $3\pi$ rest frame, while   $\psi_k$ and $\theta_k$ are the azimuthal and polar angles, respectively, of the $\pi^i$ in the $\pi^i\pi^j$ rest frame ({\it i.e.} the isobar rest frame).  For the three possible coupling schemes,  the  corresponding  Euler rotations,  $r_i$, $i=\pm,0$, are related to each other by 
   \begin{equation} 
   r_0 = r_+ r(0,\chi_+,0) = r_- r^{-1}(0,\chi_-,0),
   \end{equation} 
   where $\chi_+ (\chi_-)$ is the angle between $\pi^+$ ($\pi^-$)  and $\pi^0$ in the $3\pi$ rest  frame. This enables us to write $T$ in terms of $D(r_0)$ alone: 
      \begin{eqnarray}
    & & T_\lambda = \sum_{\mu,\nu=\pm,0}D^{1*}_{\lambda,\nu}(r_0) 
  \left[ d^1_{\nu,0}(\theta_0) \delta_{\nu\mu} F_\mu(s_{+-}) +  \right. \nonumber \\
 &&  + \left. d^1_{\mu\nu}(\chi_+)  d^1_{\mu,0}(\theta_+) F_\mu(s_{-0}) + 
  d^{1}_{\nu\mu}(\chi_-) d^1_{\mu,0}(\theta_-) F_\mu(s_{0+}) \right]. \nonumber\\
\end{eqnarray} 
 The helicity  amplitudes, $F_\mu$, are linear combinations of the  $L-S$ coupling, isospin-$I$ 
amplitudes, $F^J_{ILS}$~\cite{Ascoli:1975mn}.  In the case considered here with $I=L=S=1$, only a single  amplitude, $F^1_{111}$, contributes, and 
\begin{equation} 
F_\mu(s_{ij}) = -\frac{1}{\sqrt{6}}  \frac{3}{4\pi} \langle 1\mu| 1,\mu;1,0\rangle F^1_{111}(s_{ij}), 
\end{equation} 
which implies $F_0 = 0$ and $F_1 = -F_{-1}$. Finally,  comparing with Eq.(\ref{Dalitz}),
in the isobar model we obtain 
\begin{widetext} 
\begin{eqnarray} 
 F(s_{+-},s_{0+},s_{-0}) & = & -\frac{1}{\sqrt{6}}\frac{3}{4\pi} 
 \sum_{\nu=\pm}  (\delta_{\nu,1} + \delta_{\nu,-1}) d^1_{1,0}(\theta_0) F_1(s_{+-})
+ ( d^1_{1,\nu}(\chi_-) + d^1_{-1\nu}(\chi_-) ) d^1_{1,0}(\theta_-) F_1(s_{0+})  \nonumber \\
&  + &  (d^1_{1,\nu}(\chi_+) + d^1_{-1,\nu}(\chi_+)  ) d^1_{1,0}(\theta_+)F_1(s_{-0}).
  \label{full1}
\end{eqnarray}
\end{widetext} 

\subsection{Unitarity constraints on the isobar amplitudes} 

Writing the $J/\psi$ decay amplitude as an analytical function of the channel sub-energy, $s_{jk}$, one finds 
\begin{eqnarray} 
&&\langle (ij)k, out | J/\psi,in\rangle - \langle (ij)k,in|J/\psi,in\rangle =  (2\pi)^4 i\times   \nonumber \\
 & &  \times \sum_{i'j' }  \delta^4(p_i + p_j - p'_i - p'_j) t^*(ij;i'j')
 \langle (i'j') k,out | J/\psi,in\rangle,  \nonumber \\ \label{u1}
 \end{eqnarray}
 where  $t(ij;i'j')$ is the scattering amplitude between the incoming $|ij,in\rangle$ and the outgoing 
 $|i'j',out\rangle$ state. The two matrix elements on the {\it l.h.s.} give the $J/\psi$ decay amplitude evaluated at  $s_{ij} + i\epsilon$  and $s_{ij} - i\epsilon$, respectively. Similarly, discontinuities 
   across  the other two sub-channel energies can be considered.  However,  because of the symmetry of the isobar amplitude under permutation of the three pions, they all lead to the same unitarity relation.  
The summation over intermediate states on the {\it r.h.s.}  should include inelastic channels.  It is known that the $P$-wave $\pi\pi$ amplitude is elastic up to energies 
 $\sim1.4\mbox{ GeV}$, with the $K{\bar K}$ channel effectively saturating inelasticity above this energy, at least up to $\sim 1.9\mbox{ GeV}$ where data is available. Thus, using a single $K{\bar K}$ intermediate channel, Eq.(\ref{u1}) leads to 
     \begin{eqnarray}
  Im \hat F_\pi(s+i\epsilon)  &= &  \hat t^*_{\pi\pi}(s)  \hat \rho_\pi(s)  \hat F_\pi(s)  
  \theta(s - 4m_\pi^2)  \nonumber \\
 & + &  \hat t^*_{\pi K}(s) \hat \rho_K(s) \hat F_K(s) \theta(s - 4m_K^2). \nonumber \\
   \label{rhs1}
\end{eqnarray} 
As discussed in Section~\ref{sec:P}, this is  an approximate relation, which ignores contributions to the {\it r.h.s.} from  re-scattering between a pion from the isobar and the spectator pion. 
In Eq.(\ref{rhs1}), the helicity-$1$ isobar amplitude, $F_1$ from the {\it r.h.s.} of Eq.(\ref{full1}), is 
  denoted by  $F_\pi(s)$ to distinguish it from the corresponding helicity-$1$ amplitude 
   for production of $K{\bar K}$ $P$-wave pair in $J/\psi \to (K{\bar K})_P\pi $, which we denote by $F_K(s)$.  
Furthermore  we define $\hat F_\alpha(s)$ ($\alpha = \pi,K$) 
   as the reduced isobar amplitude, {\it i.e.} the amplitude with the angular momentum barrier factors 
   \begin{eqnarray} 
   & & 2q_\alpha(s) \equiv  \sqrt{s - s_\alpha},  \;  s_\alpha = 4m_\alpha^2, \nonumber \\
   && 2p(s)   = \sqrt{ \frac{  (M^2 - (\sqrt{s}+m_\pi)^2)(M^2 - (\sqrt{s} + m_\pi)^2) }{M^2}},  
    \nonumber \\
   \end{eqnarray}
    removed,  so that $\hat F_\alpha \equiv  F_\alpha/(2q_\alpha 2p)$.  
Here $q_\alpha$ is the relative momentum between the pions ($\alpha=\pi$) or kaons  ($\alpha = K$) 
     in the isobar rest frame, and $p$ is the break-up momentum of the $J/\psi$ (mass $M$) into 
      an  isobar of mass $\sqrt{s}$ and the spectator pion. 
     In addition,  $t_{\pi\pi}$  ($t_{K\bar K}$)  is the elastic, isospin-$1$  $\pi\pi$ ($K{\bar K}$)  $P$-wave amplitude, and $t_{\pi K}$ is the $P$-wave transition amplitude for  $K{\bar K} \to \pi\pi$. 
  Similarly, $\hat t_{\alpha\beta}$ are defined as the scattering amplitudes without the barrier factors, {\it i.e.} 
  $\hat t_{\alpha\beta} \equiv t_{\alpha\beta}/(4q_\alpha q_\beta)$.  
In terms of the $P$-wave phase shifts, $\delta_\pi$ and $\delta_K$,  and the inelasticity, $\eta$, these  amplitudes are given by, 
   \begin{eqnarray} 
& &  t_{\pi\pi} = \frac{\eta e^{2i\delta_\pi} - 1}{2i \rho_\pi },  t_{K {\bar K} } = \frac{\eta e^{2i\delta_K} - 1}{2i \rho_K},   \nonumber \\
 & & t_{\pi K} = t_{K \pi}  = \frac{\sqrt{1 - \eta^2} e^{i (\delta_\pi + \delta_K)} }{2\sqrt{\rho_\pi \rho_K}}. 
\label{ts} 
 \end{eqnarray} 
where the phase space factors are given by $\rho_\alpha(s) = \sqrt{1 - s_\alpha/s}$ 
and the $\hat \rho_\alpha$ in Eq.(\ref{rhs1}) are defined as $\hat \rho_\alpha(s) \equiv  4 q_\alpha^2(s) \rho_\alpha(s) = (s - s_\alpha) \rho_\alpha(s)$. 
Similarly one finds 
     \begin{eqnarray}
Im  \hat F_K(s+i\epsilon)   &= &  \hat t^*_{K \pi}(s) \hat \rho_\pi(s)  \hat F_\pi(s)
 \theta(s - 4m_\pi^2)  \nonumber \\
 & + &  \hat t^*_{K K}(s) \hat \rho_K(s) \hat F_K(s)  \theta(s - 4m_K^2). \nonumber \\ \label{rhs2}
\end{eqnarray}  
  In the isobar approximation the form factors $F_\pi$ and $F_K$ are real analytical functions ($\hat F_\alpha(s^*) = \hat F^*_\alpha(s)$)  of a single sub-channel energy and thus have only the unitary cuts and satisfy 
  \begin{equation} 
 \hat F_\alpha (s) = \frac{1}{\pi} \int_{s_\pi}^\infty \frac{ Im \hat F_\alpha(s') }{s' - s}  ds' . \label{cau} 
 \end{equation} 
 With $Im \hat F_\alpha$ given by Eqs.(\ref{rhs1}) and~(\ref{rhs2}) the isobar form factors  become a set of two coupled 
  integral equations.  
An analytical solution can be obtained using the standard Omn\'es-Muskhelishvili approach~\cite{M,O}. To this extent one first notices that, in the two-channel $(\alpha =\pi,K)$ approximation considered here, the unitarity condition for the reduced scattering amplitudes, ${\hat t}_{\alpha\beta}$, is given by 
  \begin{equation} 
 Im \hat t_{\alpha\beta}(s+i\epsilon)  = \sum_{\gamma=\pi,K} \hat t^*_{\alpha\gamma}(s) \hat \rho_\gamma(s) \theta(s -s_\gamma) 
  \hat t_{\gamma\beta}(s).   \label{ut} 
  \end{equation} 
This implies that the right hand discontinuity relations for $\hat F_{\alpha}$ are satisfied by the functions~\cite{Pham:1976yi} 
\begin{equation} 
\hat F_\alpha(s) = \sum_{\beta=\pi,K} \hat t_{\alpha\beta}(s) P_\beta(s),  \label{prod} 
\end{equation} 
where the production amplitudes, $P_\alpha(s)$, are real for $s>0$ and  free from right hand side 
 discontinuities. 
If $\hat F_\alpha(s)$ is to be free from discontinuities for $s < 0$ then the  production amplitudes $P_\alpha(s)$  have to satisfy the  integral equation 
\begin{equation} 
P_\alpha(s)  =  \frac{1}{\pi} \int_{-\infty}^0 ds' 
 \frac{Im P_\alpha(s')}{(s' - s)}. \label{P} 
 \end{equation}
For $s<0$,  $Im P_\alpha(s)$ is obtained from the condition $Im \hat F_\alpha(s) = 0$, 
 \begin{equation} 
 Im P_\alpha(s) = \sum_{\beta,\gamma=\pi,K}[Re \hat t(s)]^{-1}_{\alpha\beta} [Im \hat t(s)]_{\beta\gamma} Re P_\gamma(s). \label{Pim}
 \end{equation} 
In general, at most one subtraction in Eq.(\ref{P})  may be needed based on the asymptotic behavior of the scattering amplitude, which is discussed below. The subtraction constants would  
 then become  fit parameters in this  unitarized   isobar  approach.

 \subsection{ $P$-wave $\pi\pi$ scattering amplitude: general properties } 
 \label{gp} 
 
 In order to solve Eqs.(\ref{cau}) and~(\ref{P}), it is convenient to separate the left ($s<0$) and right ($s>s_\pi$)  
  cut contributions to the reduced scattering amplitudes $\hat t_{\alpha\beta}(s)$. This can be done using the  "N/D" representation independently for the amplitude of each  channel~\cite{CM}, 
     \begin{equation} 
\hat t_{\alpha \beta} = \frac{N_{\alpha\beta}(s)}{D_{\alpha\beta}(s)},
\end{equation} 
with $N_{\alpha\beta} = N_{\beta\alpha}$ and $D_{\alpha\beta} = D_{\beta\alpha}$ having only the left and right hand cuts, respectively.  Then analyticity of the amplitudes in the cut $s$-plane then  leads to~\cite{FW}
\begin{equation} 
N_{\alpha\beta}(s) = \frac{1}{\pi} \int_{-\infty}^0 ds' 
\frac{ Im t_{\alpha\beta}(s') D_{\alpha\beta}(s') }{ \sqrt{(s' - s_\alpha )(s' - s_\beta )}(s' -s)} \label{N}
\end{equation} 
and 
\begin{eqnarray}
D_{\alpha\beta}(s) &  = &  1 - \frac{(s - s_0)}{\pi} \int_{s_\pi}^\infty ds' \frac{ N_{\alpha\beta}(s') R_{\alpha\beta}(s') }{(s' - s)(s' - s_0) } \nonumber \\
 &- & \Pi_{p=1}^{N_p} \frac{s - s_0}{s_{p,\alpha\beta} - s_0} \frac{\gamma_{p,\alpha\beta}}{s_{p,\alpha\beta} - s},  \label{D} 
\end{eqnarray} 
where 
\begin{eqnarray} 
&&R_{\alpha\beta}(s)  =   \frac{Im \hat t_{\alpha\beta}(s)}{|\hat t_{\alpha\beta}(s)|^2}=  \nonumber \\
&   &  = \frac{\sqrt{s - s_\alpha}\sqrt{s - s_\beta} }{ |t_{\alpha\beta}(s)|^2} 
  \sum_{\gamma =\pi,K} t^*_{\alpha\gamma}(s) \rho_\gamma(s)\theta(s - s_\gamma) t_{\gamma\beta}(s). \nonumber \\
\end{eqnarray}
We have chosen to normalize $N_{\alpha\beta}$ and $D_{\alpha\beta}$ such that $D_{\alpha\beta}(s_0)=1$ (a convenient choice that will be employed later is $s_0 = 0$). The last term in the dispersion relation for $D_{\alpha\beta}$ reflects the so called CDD ambiguity~\cite{CDD}; the unitarity relation in Eq.(\ref{ut}) does not uniquely determine  $D_{\alpha\beta}$ if $\hat t_{\alpha\alpha}$ vanishes at some $s=s_{p,\alpha\beta}$, $p = 1,\dots, N_p$. These zeros are then incorporated as poles  in $D_{\alpha\beta}$ with $\gamma_{p,\alpha\beta}$ being their residues. 
 It  is clear from Eq.(\ref{ts}) that these poles can exist only in the elastic region of $s_K> s>s_\pi$ or in the inelastic region $s>s_K$ if inelasticity happens to vanish, $\eta = 1$ (including the point at infinity).  At every CDD pole the phase of
  the elastic amplitude passes through $180^0$ or the inelastic amplitude vanishes. 
    If the residue of a CDD pole  is small then 
  $D_{\alpha\alpha}(s)$ will develop a zero on the
 unphysical sheet near the position of the pole, {\it i.e.}  produce
 a resonance. Thus, in the past it has been proposed to identify CDD poles with the elementary  quark bound states that turn into physical resonances when coupled to
the continuum channels. Indeed it has been shown that
in potential models describing, for example, the scattering of a static source with internal structure, 
  the CDD poles correspond to excitations of the target~\cite{Lee}. 
 Asymptotically, at large $s$, $t_{\alpha\beta}(s \to \infty + i\epsilon) < O(1)$, and 
  since $D_{\alpha\beta}(s \to \infty) = O(1)$ it follows from Eqs.(\ref{N}) and~(\ref{D}), that  (for $P$-wave) $N_{\alpha\beta}(s\to \infty) = O(1/s)$. The set of coupled integral equations, Eqs.(\ref{N}) and~(\ref{D}), gives the scattering   amplitudes $t_{\alpha,\beta}(s)$ for all complex $s$ in terms of the discontinuity of the scattering amplitudes on the left cut and the location of the zeros in the physical region (the CDD poles).

  The left hand cut discontinuity plays the role of the driving term, which is analogous to the potential in  nonrelativistic  Shr\"odinger theory and in general it is not known. Fortunately, 
 as is clear from Eq.(\ref{prod}), both $N_{\alpha\beta}(s)$ and the production vectors 
 $P_\alpha(s)$ are real and have no singularities in the physical region. Thus it is the 
  behavior of the $D_{\alpha\beta}(s)$ which determine the phase and any rapid variation of the isobar amplitudes $\hat F_\alpha(s)$.    We  will use Eqs.(\ref{N}) and~(\ref{D}), not as integral equations for $N$ and $D$, but instead we will use what  is known about the scattering amplitude at the boundary of the right hand cut, $\hat t_{\alpha\beta}(s+ i\epsilon)$,  with a model for the left hand cut as input to determine the denominator functions.  
 Then  Eq.(\ref{D}) can be written as an integral equation for $D$ alone 
  \begin{eqnarray}
& & D_{\alpha\beta}(s)   =  1 - \Pi_{i=p}^{N_p} \frac{s - s_0}{s_{p,\alpha\beta} - s_0} \frac{\gamma_{p,\alpha\beta}}{s_{p,\alpha\beta} - s}  \nonumber \\
- &  & 
 \frac{(s - s_0)}{\pi} \int_{s_\pi}^\infty ds' \frac{ D_{\alpha\beta}(s') e^{-i\phi_{\alpha\beta}(s')}
 \sin\phi_{\alpha\beta}(s')  }{ (s' - s)(s' - s_0) }, \nonumber \\
 \label{D1} 
\end{eqnarray} 
where $\phi_{\alpha\beta}$ is the phase of $t_{\alpha\beta} = |t_{\alpha\beta }(s)|\exp(i\phi_{\alpha\beta}(s))$,
which has an analytical solution given by 
 \begin{equation} 
 D_{\alpha\beta}(s) = \Pi_{p=1}^{N_p} \left( \frac{s_0 - s_{p,\alpha\beta}}{s- s_{p,\alpha\beta}} \right)
  \Pi _{q=1}^{N_q} \left( \frac{s- s_{q,\alpha\beta}}{s_0 - s_{q,\alpha\beta}}\right)
   \Omega_{\alpha\beta}(s).   \label{OM}
 \end{equation} 
The first (second)  factor  gives the contribution from the CDD poles (zeros)
 and  $\Omega$ is the 
 Omn\'es-Muskhelishvili function, 
 \begin{equation}
 \Omega_{\alpha\beta} (s) = \exp\left( - \frac{s - s_0}{\pi} \int_{s_\pi}^\infty ds' 
 \frac{\phi_{\alpha,\beta}(s')}{(s' - s)(s' - s_0)} \right). \label{OM1} 
 \end{equation} 
Phase shifts $\delta_\alpha$  are determined up to an integer  multiple of $\pi$ and the phase of the amplitude $\phi_{\alpha\beta}$ is determined modulo $2\pi$.  It is customary to remove this ambiguity by setting all phase shifts to zero at elastic thresholds, {\it i.e.} $\delta_\alpha(4m_\alpha^2) = 0 $. 
This  condition is at the origin of  zeros of $D_{\alpha\beta}$ being  explicit in Eq.(\ref{OM}). 
 With $\phi_{\alpha\beta}(4 m_\pi^2) = 0$  and the asymptotic behavior, $D_{\alpha\beta}(s\to \infty) = O(1)$, the number of zeros, $N_q$, and CDD poles, 
$N_p$, are related by  
\begin{equation}
\phi_{\alpha\beta}(\infty)  = \pi ( N_p - N_q). \label{lev}
\end{equation}

\subsection{Analytical model for the $P$-wave amplitude} 

    \begin{figure}[ht]
\begin{center}
\includegraphics[width=3.5 in,angle=00]{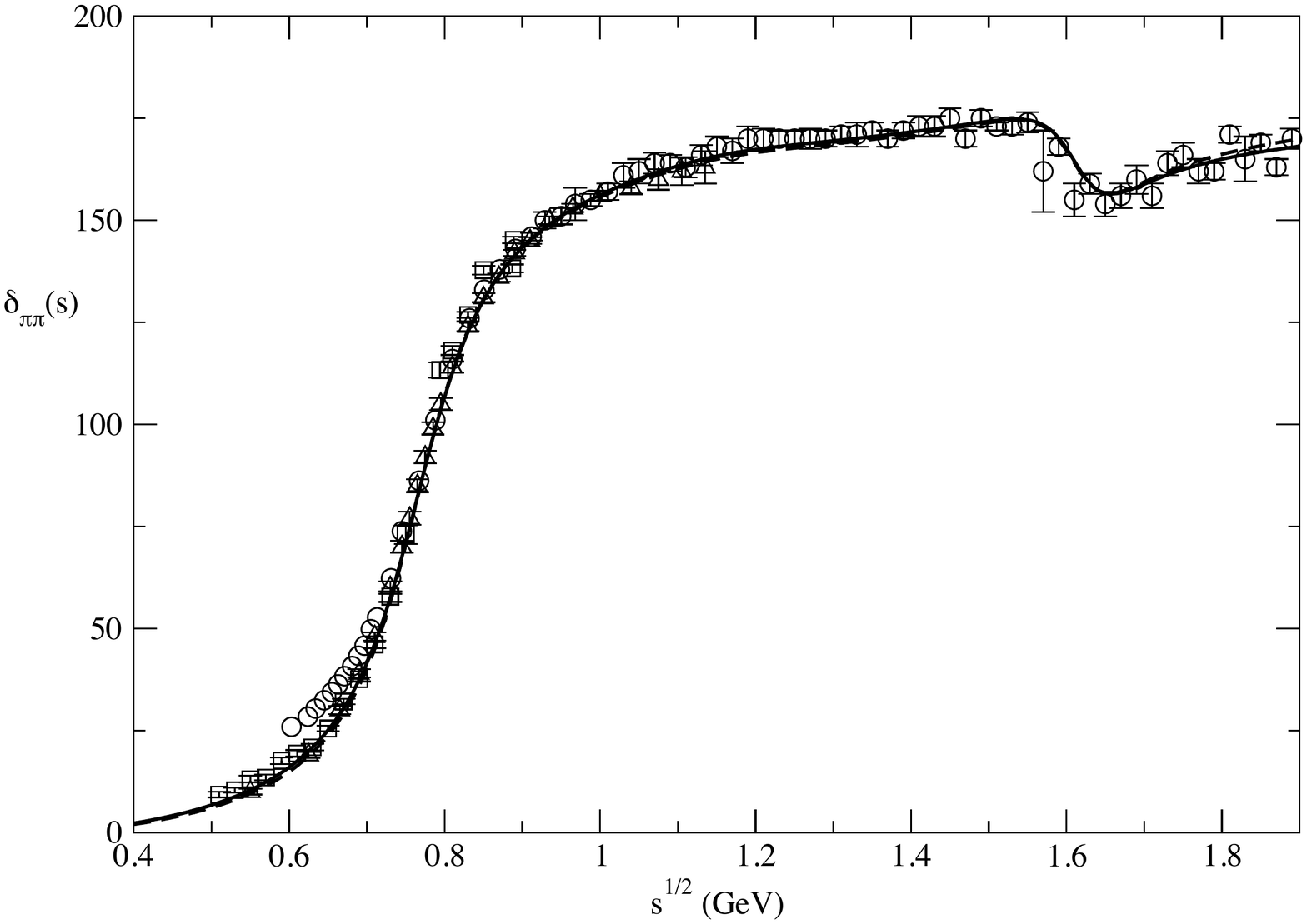}  
\includegraphics[width=3.5 in,angle=00]{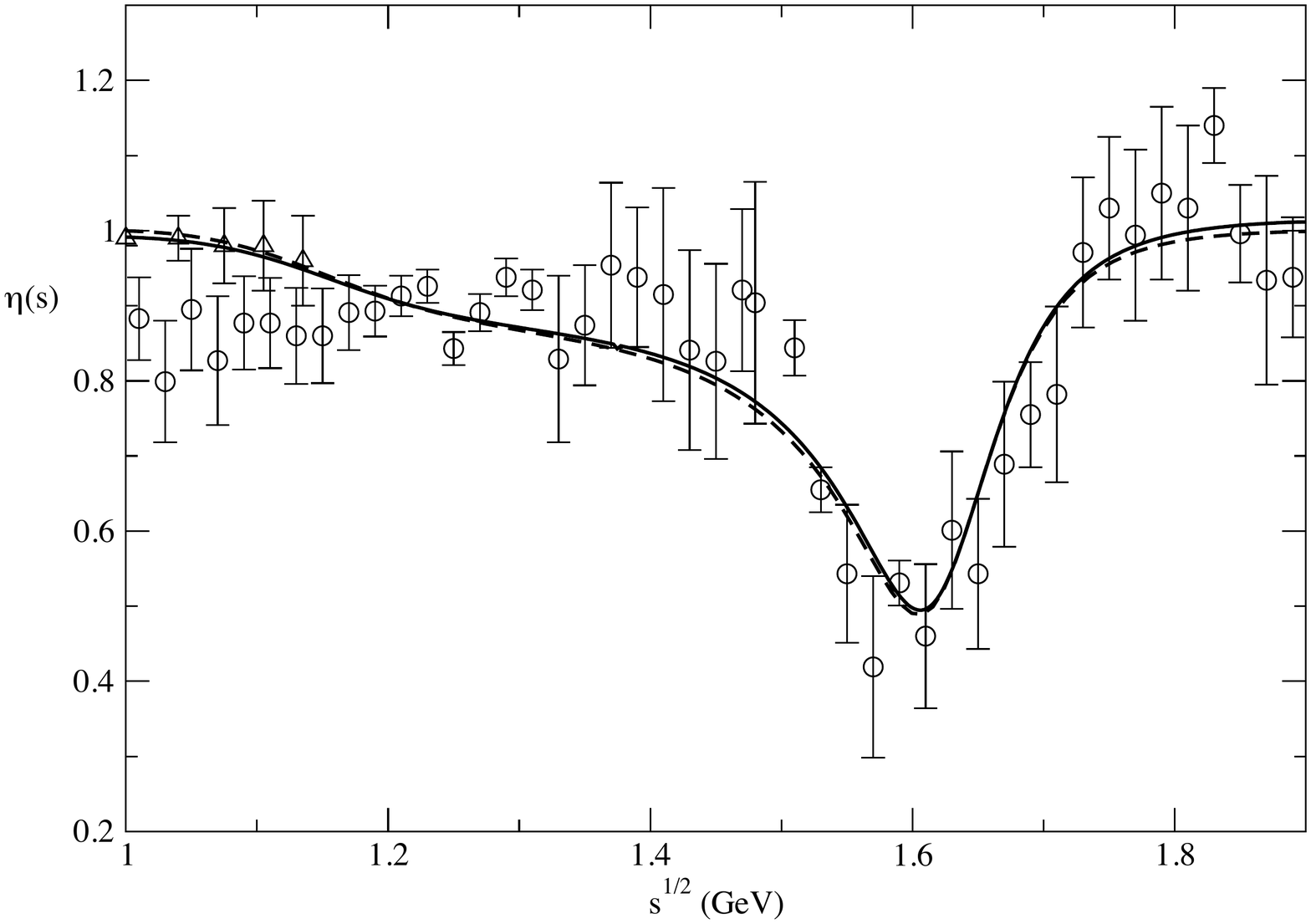}  
\caption{Phase shift~(top) and inelasticity~(bottom) of the $P$-wave $\pi \pi$ amplitude. Data is taken from 
 \cite{Hyams:1973} (circles) ,\cite{Protopopescu:1973} (triangles) , and \cite{Estabrooks:1974} (squares). 
   The solid line  is the result of the fit  to $\delta_\pi$ and $\eta$  with the analytical $K$-matrix representation described  in the text. The dashed line is the result of the extended parametrization described in Section~\ref{sec:modifedK}. 
\label{fig:phase}}
\end{center}
\end{figure}

If the left hand cut discontinuity  of $\hat t_{\alpha\beta}(s)$ were known, then the whole amplitude could be reconstructed using the $N/D$ method discussed above and the production vectors $P_\alpha(s)$  could be computed from Eq.(\ref{P}).  Unfortunately, to the best of our knowledge, 
 only in the case of  $\hat t_{\pi\pi}$  is the  left hand cut fairly well known~\cite{Tryon:1974tq}.  Thus, one needs a model to incorporate the contribution    from the $K{\bar K}$ channel.  One might as well then construct a model that leads to a simple  solution of  the integral equation in  Eq.(\ref{P}). This is indeed the case if one 
   uses the analytical $K$-matrix representation 
   with the typical choice of  the $K$-matrix  parametrized in terms of simple poles. Then the singularity of the scattering  amplitude for $s<0$ is also given by poles and this in turn allows one to solve  Eq.(\ref{P}) by algebraic methods.    We fix the parameters of the $2\times 2$   $K$-matrix so as to reproduce the $P$-wave $\pi\pi$ data from \cite{Hyams:1973,Protopopescu:1973,Estabrooks:1974} (Fig.~\ref{fig:phase});  $\delta_\pi$ and $\eta$ are input parameters, and the model will  give  a prediction for $\delta_K$. The $K$-matrix parametrization was already used by Haymes~{\it at el.} to interpret their data from \cite{Hyams:1973}. Unfortunately, instead of using Eq.(\ref{ut}), the unitarity condition  employed in \cite{Hyams:1973} was 
    \begin{equation} 
 Im \hat t_{\alpha\beta}(s+i\epsilon)  = \sqrt{s} \sum_{\gamma=\pi,K} \hat t^*_{\alpha\gamma}(s) \hat  \rho_\gamma(s) \theta(s -s_\gamma) 
  \hat t_{\gamma\beta}(s).    \label{uth}     
\end{equation} 
This implies 
\begin{equation} 
Im [ \hat t^{-1}(s)]_{\alpha\beta}(s)  =  -(s - s_\alpha)
\sqrt{s - s_\alpha} \delta_{\alpha\beta} ,
\end{equation} 
and the $K$-matrix representation becomes
\begin{equation} 
[\hat t^{-1}(s)]_{\alpha\beta} = [K^{-1}(s)]_{\alpha\beta} + \delta_{\alpha\beta} (s - s_\alpha) \sqrt{s_\alpha - s}. 
\end{equation} 
In contrast, the correct unitarity relation in Eq.(\ref{ut}) gives 
\begin{equation} 
Im [ \hat t^{-1}(s)]_{\alpha\beta}(s)  =  -(s - s_\alpha) 
\sqrt{\left( 1 - \frac{s_\alpha}{s}\right)} \delta_{\alpha\beta}, 
\end{equation} 
which leads to
\begin{equation} 
[ \hat t^{-1}(s)]_{\alpha\beta} = [K^{-1}(s)]_{\alpha\beta} + \delta_{\alpha\beta}  (s - s_\alpha)I_\alpha(s),   \label{tk}
\end{equation} 
where 
\begin{equation} 
I_\alpha(s) = I_\alpha(0) - \frac{s}{\pi} \int_{s_\alpha}^\infty ds'   \sqrt{1 - \frac{s_\alpha}{s'}} \frac{1}{ (s' - s)s'}.  
\end{equation} 
A convenient choice for the subtraction constant, $I_\alpha(0)$, is to take $Re I_\alpha(M^2_\rho) = 0$. Then one of the poles of  $K_{\pi\pi}$  corresponds to the Breit-Wigner mass squared, $M^2_\rho=(0.77\mbox{ GeV})^2$,  of  the $\rho$ meson. Using the general two-pole parametrization of the $K$ matrix, 
 \begin{eqnarray}
& & K_{\pi\pi} =  \frac{ \alpha_\pi^2 }{M_{\rho}^{2}-s}   +\frac{\beta_{\pi}^{2} }{s_{2}-s} + \gamma_{\pi \pi}, \; 
K_{KK} =  \frac{ \beta^{2}_{K}}{s_{2}-s} + \gamma_{KK} \nonumber \\ 
& & K_{\pi K}  = K_{K\pi} =  \frac{\beta_{\pi} \beta_{K}}{s_{2}-s} + \gamma_{\pi K} ,
 \end{eqnarray}
 where $\alpha_\pi^2 = \Gamma_{\rho}M^2_{\rho}/(M_\rho^2 - s_\pi)^{3/2}$.  By 
 fitting  the $P$-wave $\pi\pi$ phase shift, $\delta_\pi$, and the inelasticity, $\eta$, we find 
  $\Gamma_{\rho} = 0.140 \mbox{ GeV}$, and 
\begin{eqnarray} 
&&   \sqrt{ s_{2}}  = 1.4708 \mbox{ GeV},  \ \ \ \  \beta_{\pi} = 0.199, \ \ \ \  \beta_{K} =0.899,
 \nonumber  \\  && \gamma_{\pi \pi} = 5.62\times 10^{-2}, \ \ \ \ \ \gamma_{\pi K} = 0.104, \ \ \ \ \ \gamma_{KK} = 1.525, \nonumber \\
 \end{eqnarray}
  with the $\gamma$'s in units of $\mbox{ GeV}^{-2}$. The comparison of the phase shift and the inelasticity obtained with this parametrization with the data is shown in Fig.~\ref{fig:phase}.

\begin{figure}[hh]
\begin{center}
\includegraphics[width=3.5 in,angle=00]{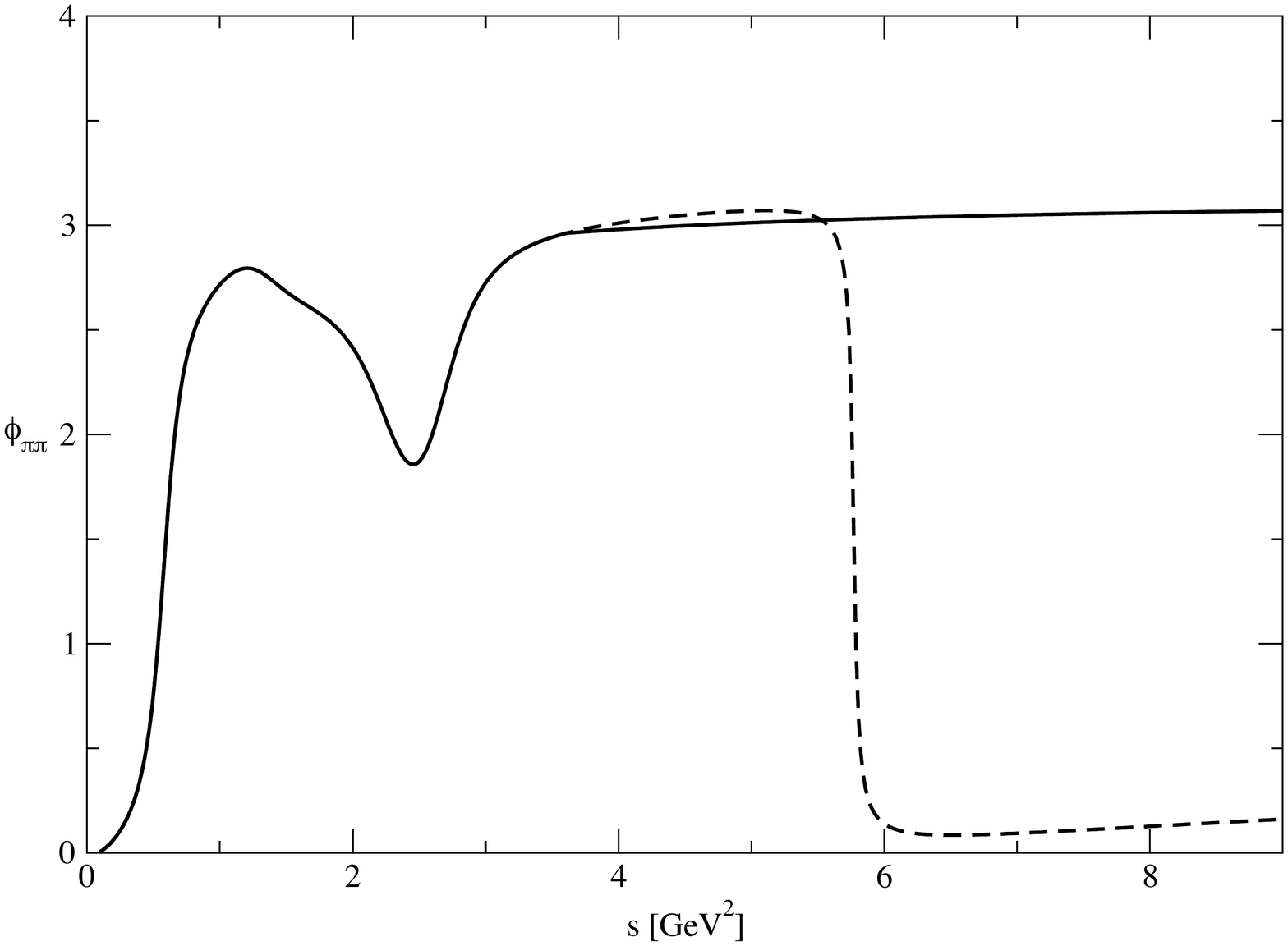}  
\includegraphics[width=3.5 in,angle=00]{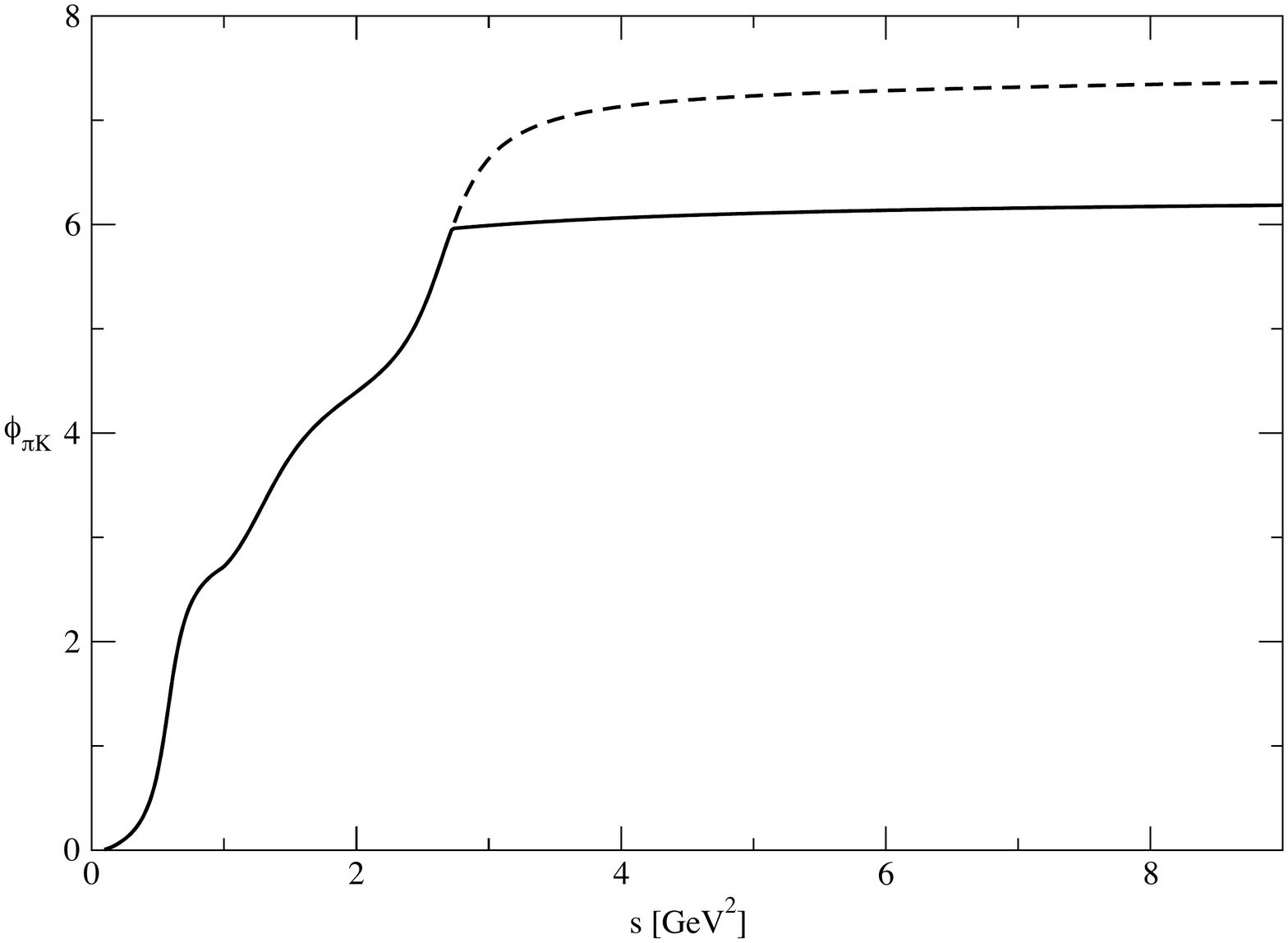}  
\caption{ Phase of the $\pi\pi$ (upper) and $\pi K$ (lower) amplitude. 
 The dashed line is the result of the $K$ matrix parametrization from Eq.(\ref{KND}). The solid line is
  from the modified  $K$ ~matrix parametrization discussed in Sec.~\ref{sec:modifedK} 
  \label{fig:phipp}}
\end{center}
\end{figure}

  Since the $K$ matrix representation 
   of  Eq.(\ref{tk}) satisfies all of the properties of the scattering amplitude discussed in Sec.\ref{gp} it is possible to write  $t_{\alpha\beta}$ in the "N/D" representation. We find, choosing to 
 normalize $D_{\alpha\beta}(s)$ at $s_0 =0$,  
 \begin{widetext} 
 \begin{eqnarray} 
&& N_{\pi\pi}(s) = \lambda_{\pi\pi} 
\frac{s - z_{\pi\pi}}{(s - s_{L,1})(s - s_{L,2})}, \; D_{\pi\pi}(s) = \exp\left(-\frac{s}{\pi} 
\int_{s_\pi}ds' \frac{\phi_{\pi\pi}(s')}{s' (s' - s)}\right), \nonumber \\
& & N_{\pi K}(s) =
\frac{ \lambda_{\pi K} }{(s - s_{L,1})(s - s_{L,2})}, \; D_{\pi K}(s) = \frac{ s_{1,\pi K} s_{2,\pi K} }{ (s -s_{1,\pi K} )(s - s_{2,\pi K})}
\exp\left(-\frac{s}{\pi} \int_{s_\pi}ds' \frac{\phi_{\pi K}(s')}{s' (s' - s)}\right), \nonumber \\
& & N_{KK}(s) =  \lambda_{KK} 
\frac{s - z_{KK}}{(s - s_{L,1})(s - s_{L,2})}, \; D_{KK}(s) = \exp\left(-\frac{s}{\pi} \int_{s_\pi}ds' 
\frac{\phi_{KK}(s')}{s' (s' - s)}\right), \nonumber \\ \label{KND}
\end{eqnarray}
\end{widetext}
with $\lambda_{\pi\pi} = 5.649$, $\lambda_{KK} = 2.271 $ and $\lambda_{\pi K} = 3.048\mbox{ GeV}^2$.  
Indeed, as discussed above, the left hand cut is reduced to two poles at $s_{L,1} = -13.87\mbox{GeV}^2$ and $s_{L,2} = -0.787\mbox{ GeV}^2$, respectively. There are also  first order 
  zeros  in $N_{\alpha\beta}$ at 
 $z_{\pi\pi} = -0.867\mbox{ GeV}^2$ and  $z_{KK} = -13.78\mbox{GeV}^2$. 
  The numerator functions for the  elastic amplitudes $\pi\pi$ and $K{\bar K}$ 
 are $O(1/s)$, and for the inelastic amplitudes they are  super-convergent, {\it i.e.} 
 $O(1/s^2)$.  Asymptotically, as shown in Fig.~\ref{fig:phipp},    
 $\phi_{\pi\pi}(s \to \infty) = O(1/\log(s))$ and $\delta_\pi$ stays below $180^0$, so there is no CDD pole in the $\pi\pi$ channel, which is consistent with the Levinson theorem ({\it cf.}~Eq.(\ref{lev})).   The same is true for the $K{\bar K}$ channel. 
 Above the $K{\bar K}$ threshold  the phase of the inelastic amplitude $\phi_{\pi K}$ is given by $\phi_{\pi K} = \delta_\pi + \delta_K$ and from the $K$ matrix we find that  asymptotically $\phi_{\pi K}(\infty) = 2\pi$, which results 
  in two CDD poles -- one at the $\rho$ mass, $s_{1,\pi K} = M_\rho^2$, and the other at   $s_{2,\pi K} =  s_2 + \beta_\pi\beta_K/\gamma_{\pi K} = 3.884\mbox{ GeV}^2$.

   \begin{figure}[hh]
\begin{center}
\includegraphics[width=3. in,angle=00]{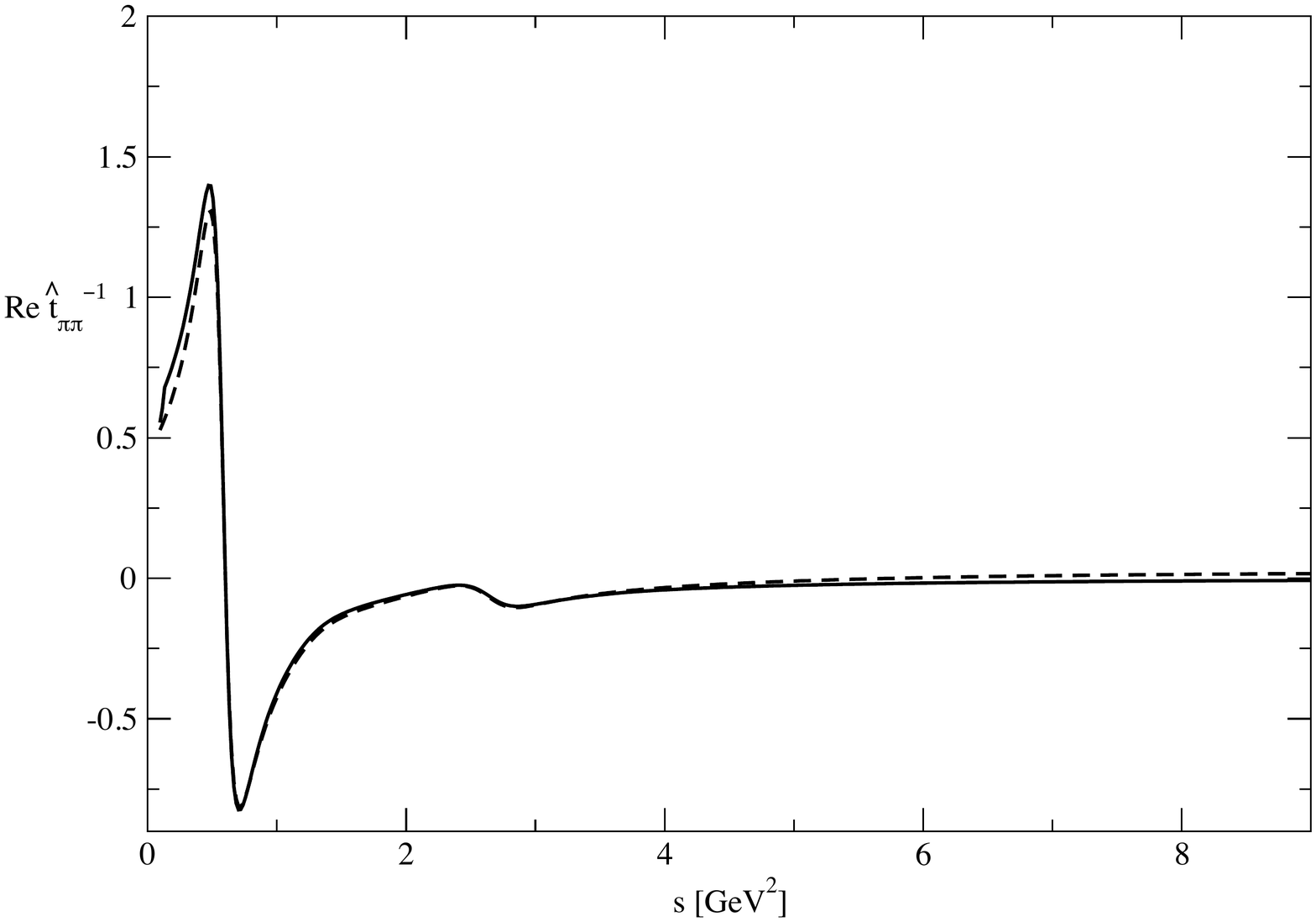}  
\includegraphics[width=3. in,angle=00]{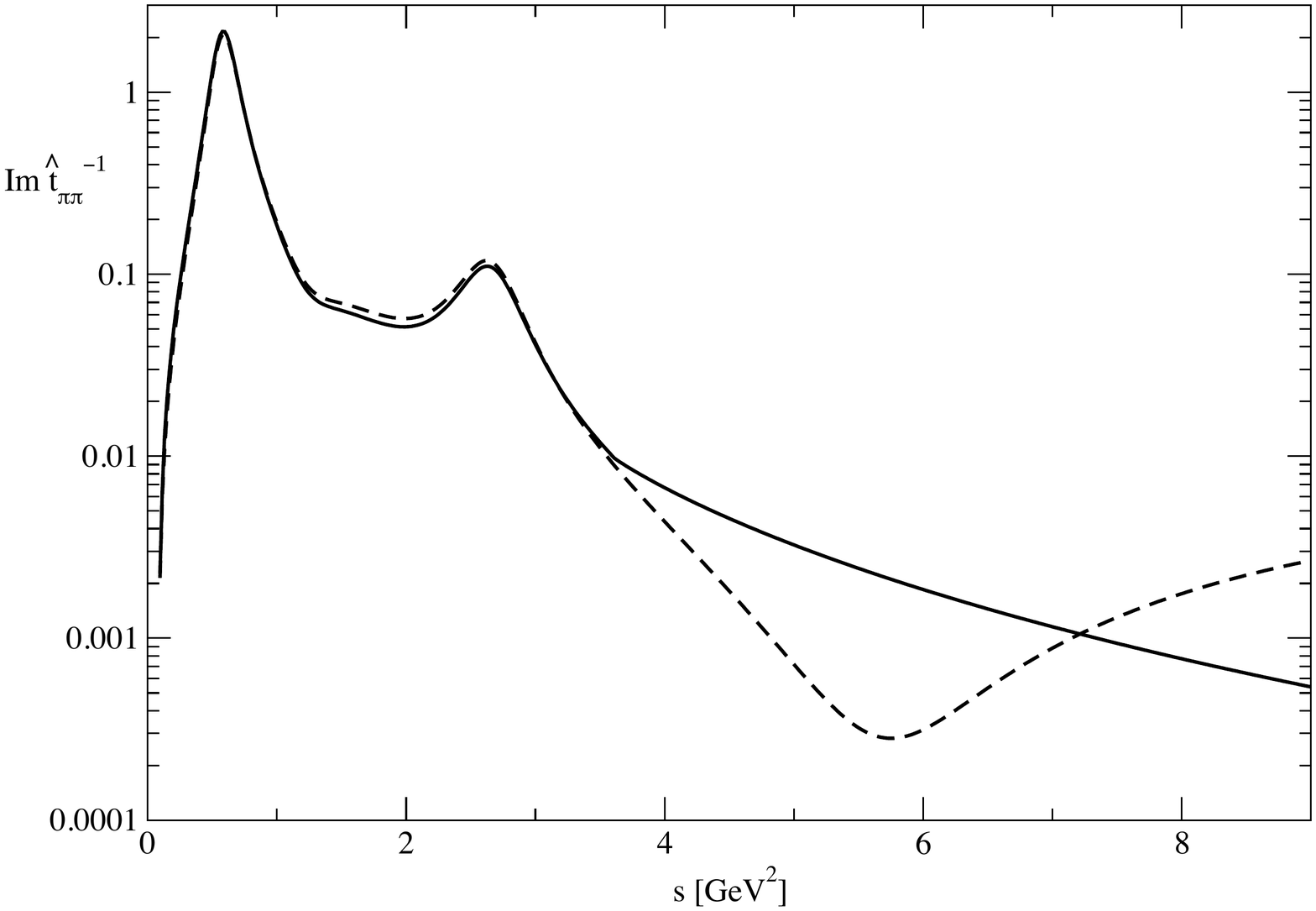}  
\caption{   Real (upper) and imaginary (lower) part of $\hat t_{\pi\pi}$. 
 The dashed line corresponds to the $K$-matrix solution of E.q~\ref{KND}, and the solid line is the 
  modified $K$-matrix solution, $\hat t^{new}_{\pi\pi}$, discussed in Sec.~\ref{sec:modifedK}, 
  Eq.(\ref{tnew11}). 
 \label{fig:tpipi}  } 
\end{center}
\end{figure}

   \begin{figure}[hh]
\begin{center}
\includegraphics[width=3. in,angle=00]{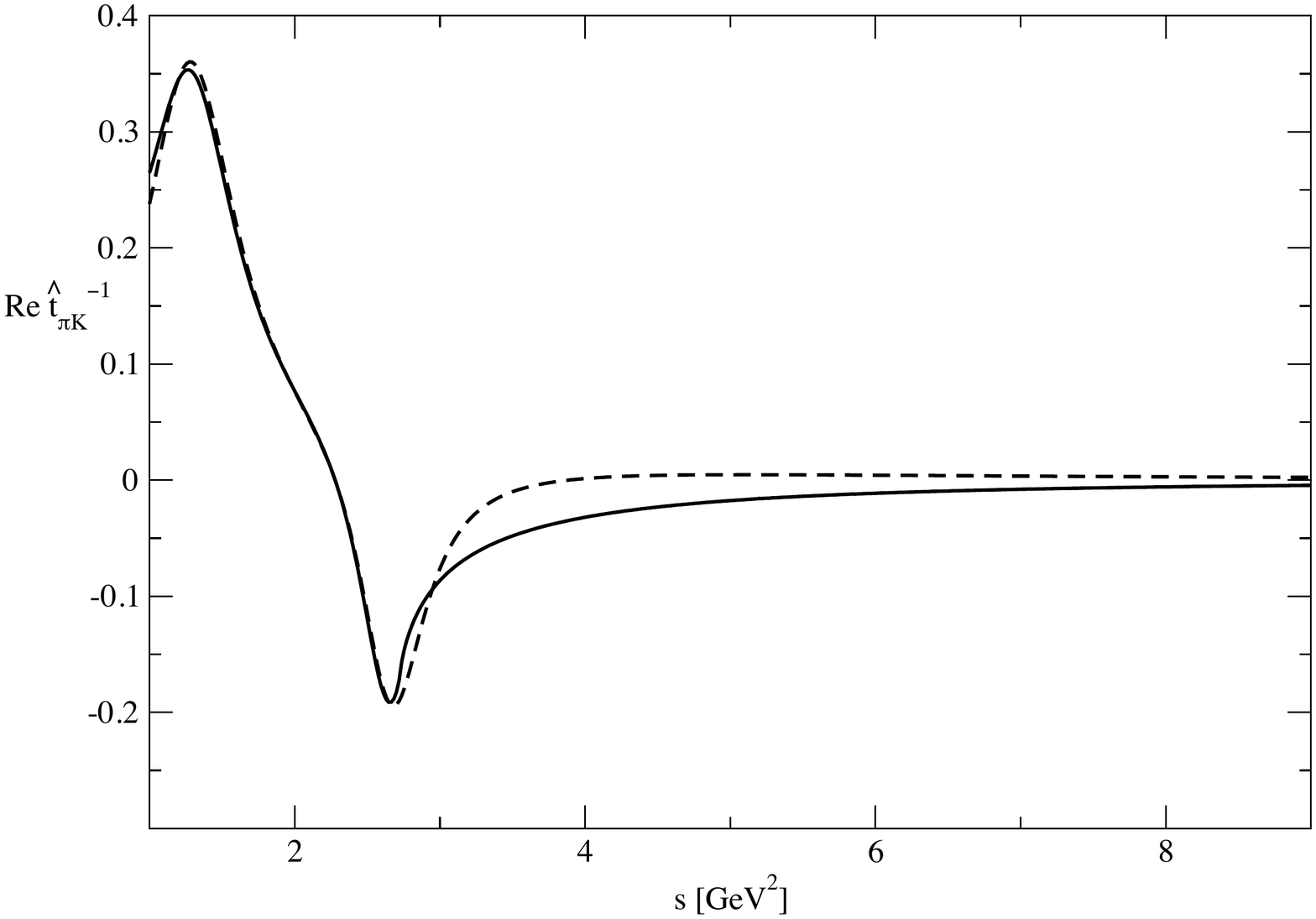}  
\includegraphics[width=3. in,angle=00]{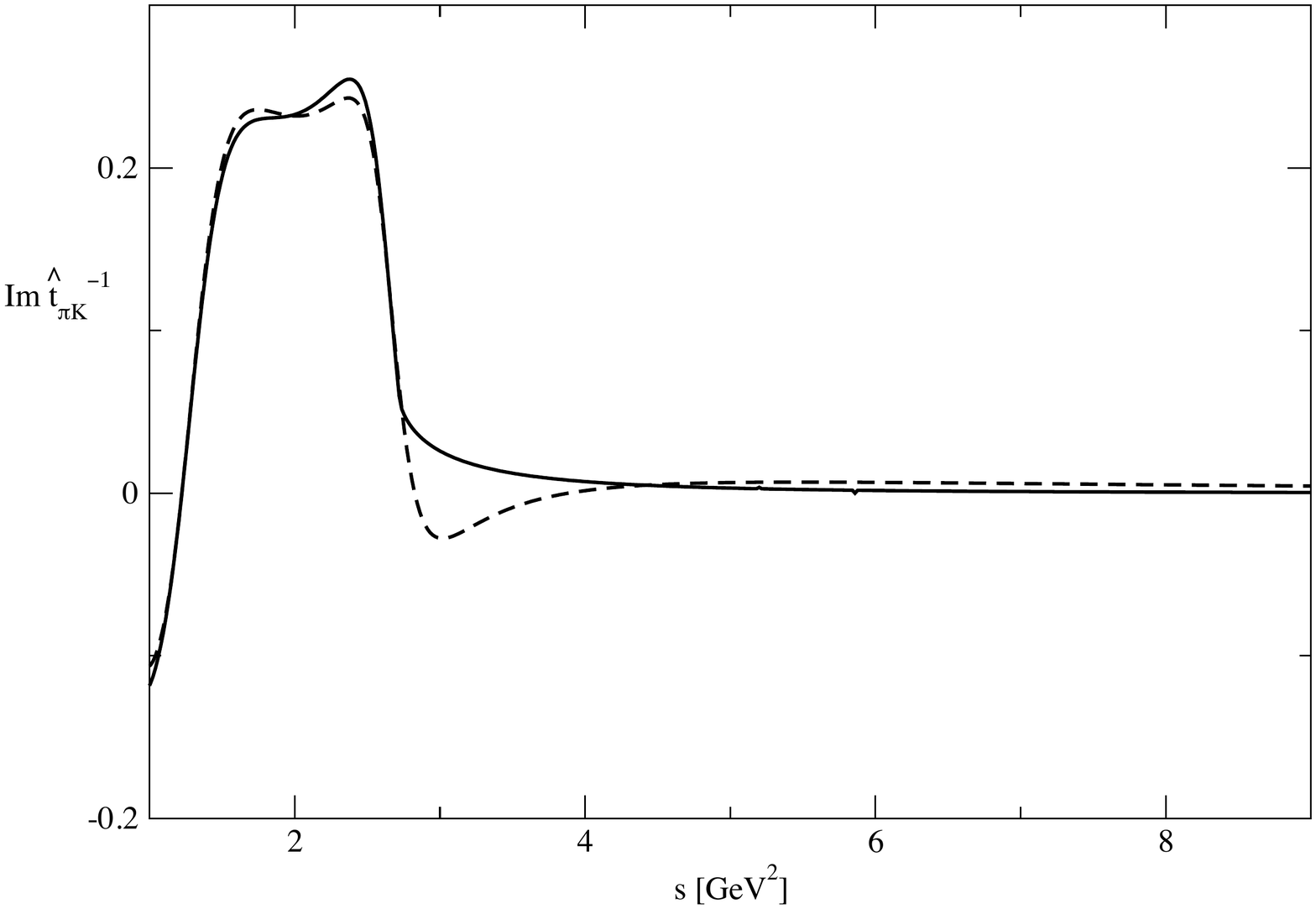}  
\caption{  Same as Fig.~\ref{fig:tpipi}  for $\hat t_{\pi K}$ from Eq.(\ref{KND}) (dashed) and 
$\hat t^{new}_{\pi K}$ from Eq.(\ref{tnew12}).  
\label{fig:tpik}}
\end{center}
\end{figure}

    \begin{figure}[hh]
\begin{center}
\includegraphics[width=3. in,angle=00]{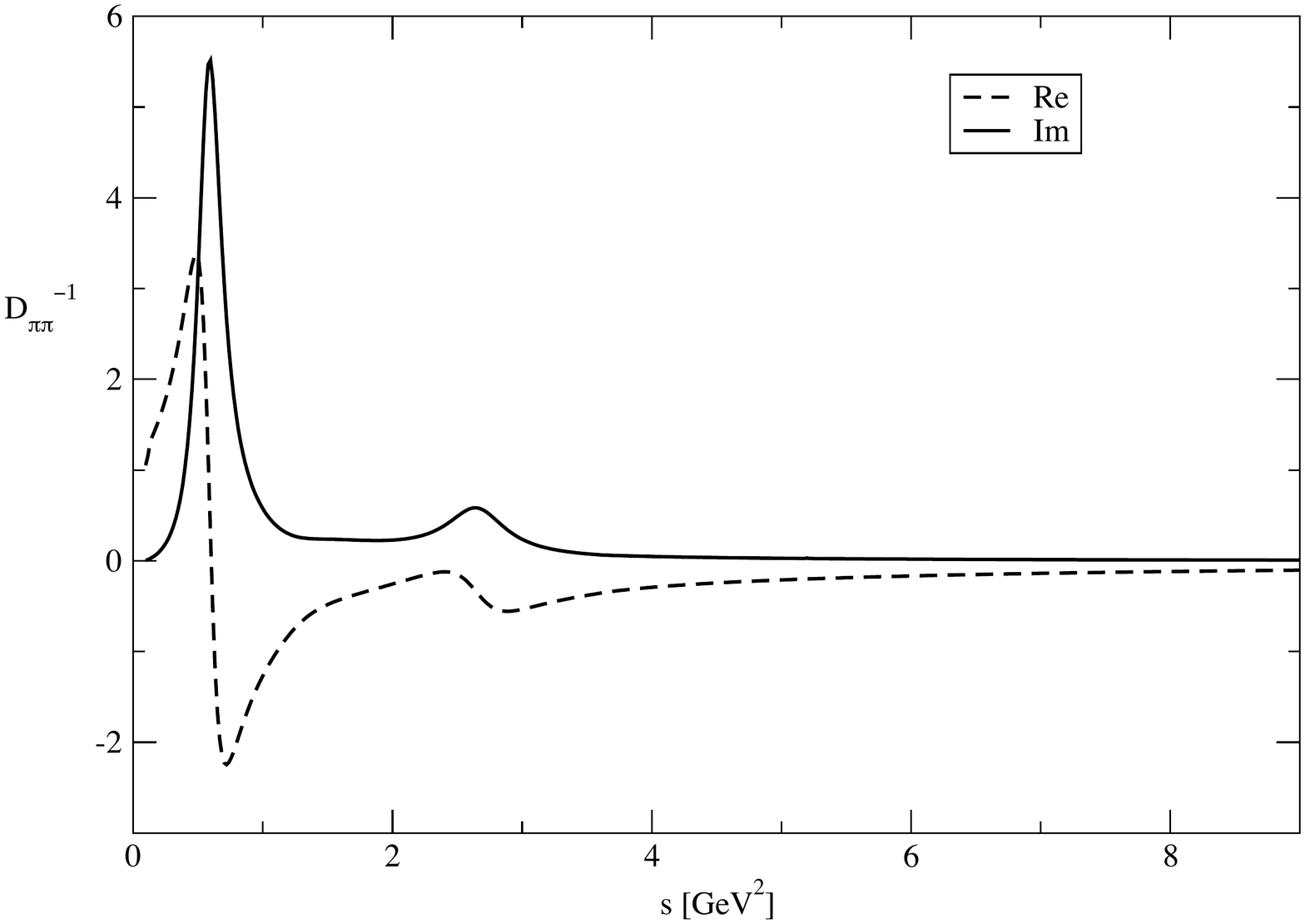}  
\includegraphics[width=3. in,angle=00]{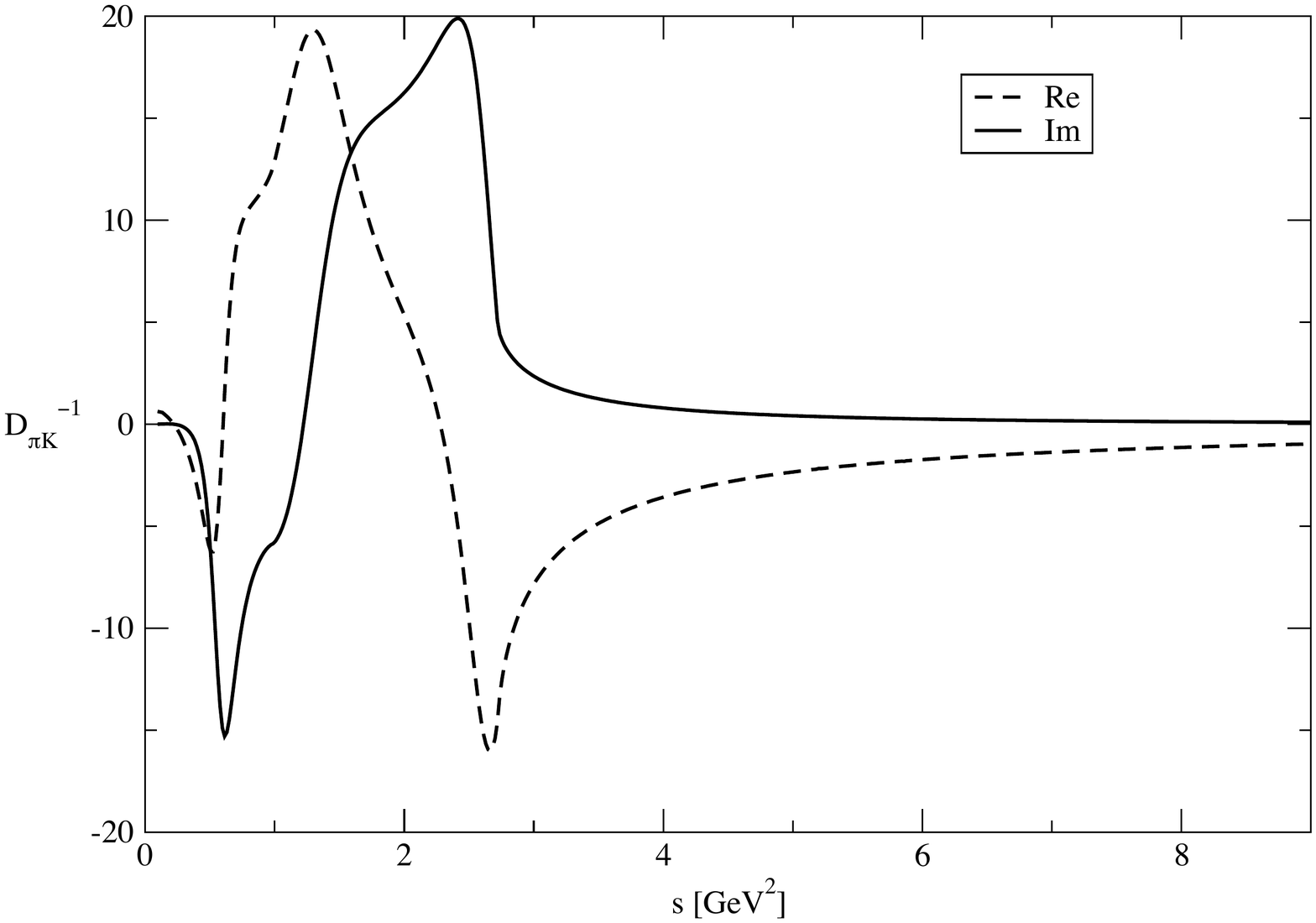}  
\caption{ Real (dashed) and imaginary (solid) part of the inverse of  $D^{new}_{\pi\pi}(s)$ (upper) and 
 $D^{new}_{\pi K}(s)$ (lower)  used in  the computation of the isobar form factor, {\it cf.} Eq.(\ref{f1}).
 \label{fig:Ds}}
\end{center}
\end{figure}

    Having an analytical representation for the scattering amplitude enables one to identify the 
 resonance content by studying the singularities of $\hat t_{\pi\pi}(s)$ for $s$  continued 
  through the unitarity cuts away from the physical sheet. 
  If we define the unphysical  sheet~II  as the one obtained by continuing $s$ from above (crossing) the cut $s_\pi < s < s_K$, and sheet~III for $s$ continued through the $s > s_K$ cut, 
      then we find four poles whose location is given in Table \ref{pole}. 
 The $\rho$ pole is clearly seen as well as the  excited $\rho'$ resonance at  $1600\mbox{ MeV}$ that couples primarily to the $K\bar K$ channel. 
     The pole on sheet~III at $1.1409-i 0.1675$ GeV is most sensitive to the inelasticity of the $K\bar{K}$ channel. If we turn off the $K\bar{K}$ channel  this pole goes to infinity while the positions of the other two remain relatively unchanged. 
     
  \begin{table}[htdp]
\caption{ Physical poles ($\sqrt{s}$ in GeV) on sheets~II and~III. \label{pole}}
\begin{center}
\begin{tabular}{|c|c|} \hline
 II & III  \\  \hline
$0.7638-i 0.0747$ & $0.7632-i 0.0745$ \\ \hline
  & $1.1409-i 0.1675$  \\ \hline
   & $1.6306-i 0.0844$  \\ \hline
\end{tabular} 
\end{center}
\label{default}
\end{table}

\subsection{ Problems with the $K$-matrix parametrization} 
\label{sec:modifedK} 
While the $K$~matrix parametrization faithfully reproduces the $\pi\pi$ phase shift and inelasticity data  from $\pi\pi$ threshold  up to $1.9\mbox{ GeV}$,  extrapolation  beyond this range is problematic.  The rapid decrease of $\phi_{\pi\pi}$ around $s \sim 6\mbox{ GeV}^2$ seems unphysical and results in an absence of the CDD pole at infinity, {\it i.e.} $\phi_{\pi\pi}(\infty) \to 0$ instead of $\phi_{\pi\pi}(\infty) \to \pi$~\cite{De Troconiz:2001wt}.  The CDD pole at infinity in the elastic $\pi\pi$    
amplitude is expected based on the asymptotic pQCD prediction for the pion electromagnetic form factor~\cite{Lepage:1980fj}. In  the $\pi\pi \to K{\bar K}$ channel, the two CDD poles 
  at $m_\rho^2$ and $s_2 + \beta_\pi\beta_K/\gamma_{\pi K}$ are clearly an artifact  of the 
   pole  parametrization of the $K$-matrix.  A CDD pole in the inelastic channel above threshold  ({\it e.g} the pole at $s_{2,\pi K} = 3.884\mbox{ GeV}^2$)  leads to a discontinuity  in a phase shift and is unphysical. A pole between $\pi\pi$ and $K{\bar K}$ thresholds is admissible, {\it e.g.} the pole at $s_{1,\pi K} = m_\rho^2$, but its strict overlap with the $\rho$ mass is an artifact of the parametrization.  Since the  phase space available in $J/\psi$ decay extends  up to $s_{\pi\pi} \sim 9 \mbox{GeV}^2$ we need to remove these unphysical features of the $K$-matrix amplitude. We proceed as follows. 
The new $\pi\pi \to \pi\pi$ and $K{\bar K} \to \pi\pi$  amplitudes  will be denoted by 
 $\hat t^{new}_{\pi\pi}(s)$ and $\hat t^{new}_{\pi K}(s)$, respectively. 
   In the case of the $\pi\pi \to \pi\pi$ elastic amplitude, we 
  assume that it has a single CDD pole at infinity. We thus introduce an effective phase shift 
   and inelasticity that asymptotically approach $\pi$ and  $1$, respectively: 
      \begin{equation} 
   \delta_{eff}(s) = \left\{ \begin{array}{c} \delta_\pi(s), s < s_K    \\ 
     \pi + ( \delta_\pi(s_K)  - \pi) \frac{s_K}{s}, s  > s_K  \end{array},\right.  
     \end{equation} 
     \begin{equation} 
      \eta_{eff}(s) = \left\{ \begin{array}{c} \eta_\pi(s), s < s_K  \\
       1 + (\eta_\pi(s_K) - 1) \frac{s_K}{s}, s  > s_K \end{array} \right. 
       \end{equation} 
   with $\sqrt{s_K} = 1.9\mbox{ GeV}$ and  $\delta_\pi$ and $\eta_\pi$ obtained from the $K$-matrix fit  below $1.9\mbox{ GeV}$ 
    ({\it cf.} Fig.~\ref{fig:phase}).  The denominator $D^{new}_{\pi\pi}$ of the 
   effective amplitude
   \begin{equation} 
   \hat t^{new}(s) = \frac{N^{new}_{\pi\pi}(s)}{D^{new}_{\pi\pi}(s)}  \label{tnew11} 
   \end{equation} 
       is then obtained from Eq.(\ref{OM1}) with $N_q = N_p = 0$ and  phase, $\phi^{eff}_{\pi\pi}$,    
        given by (see Fig.~\ref{fig:phipp})
           \begin{equation} 
   \phi^{new}_{\pi\pi}  = \mbox{ Im}  \ln \left[ \frac{\eta^{eff} e^{2i\delta_{eff}} - 1}{ 2i\hat \rho} \right].
   \end{equation} 
 
For the numerator function $N^{new}_{\pi\pi}$, we use a simple pole approximation to the left hand cut  ($s_L < 0$) 
\begin{equation} 
N^{new}_{\pi\pi}(s) = \frac{\lambda^{new}_{\pi\pi}}{s - s_L}.  \label{Nnew11} 
\end{equation} 
In order to remove the unphysical CDD pole 
 from the $K{\bar K} \to \pi\pi$ amplitude for $D^{new}_{\pi K}(s)$ in 
\begin{equation} 
\hat t^{new}_{\pi K}(s) = \frac{N^{new}_{\pi K}(s)}{D^{new}_{\pi K}(s)}  \label{tnew12} 
\end{equation} 
 for  $\phi^{new}_{\pi K}$ in Eq.(\ref{OM1}), we use (see Fig.~\ref{fig:phipp}) 
 \begin{equation} 
\phi^{new}_{\pi K} = \left\{ \begin{array}{c}  \phi^K_{\pi K}(s), s < s_K   \\ 
 2\pi + (\phi^K_{\pi K}(s) - \pi )\frac{s_K}{s}, s > s_K 
\end{array} \right. 
   \end{equation} 
with $\sqrt{s_K}  =1.65 \mbox{ GeV}$. In this case we use the $K$-matrix fit up to a lower 
 energy of $1.65\mbox{ GeV}$ to be less sensitive to the unwanted CDD 
   pole in the $K$ matrix at $\sqrt{s_{2,\pi K}}  =   1.97$.  
 There is no 
    effect of this pole in the elastic amplitude, and thus for that case we could use the $K$-matrix parametrization all the way up to $1.9\mbox{ GeV}$ where data exists. 
 Assuming further that  $D^{new}_{\pi K}(s)$ has the same asymptotic behavior as $D^{new}_{\pi\pi}$ we add a single CDD pole at $s^{new}_{1,\pi K}$ in place of the pole at  $m_\rho^2$ between the $\pi\pi$ and $K{\bar K}$ thresholds. 
  Finally,  for the numerator function we use 
\begin{equation} 
N^{new}_{\pi K}(s) = \frac{\lambda^{new}_{\pi K}}{s - s_L}.  \label{Nnew12} 
\end{equation} 
{\it i.e} we use the same pole to represent the left hand cut as in 
  $N^{new}_{\pi\pi}$. The four  parameters $\lambda^{new}_{\pi\pi}$, 
  $\lambda^{new}_{KK}$, $s_L$, and $s^{new}_{1,\pi K}$ are determined by simultaneously fitting $\hat t^{new}_{\pi\pi}$ and $t^{new}_{\pi\pi}$ to $\pi$ and $K$ phase shifts and inelasticity in the range 
  $2m_\pi < \sqrt{s} < 1.9\mbox{ GeV}$ and $2m_K < \sqrt{s} < 1.65 \mbox{ GeV}$, respectively. 
 The comparison with the $K$-matrix solution is shown in Figs.~\ref{fig:tpipi} and~\ref{fig:tpik} and the fit yields $\lambda^{new}_{\pi\pi} = 0.750$, $\lambda^{new}_{\pi K} = 0.0477$, $s_L = -1.328\mbox{ GeV}^2$, and $s^{new}_{1,\pi K} = 0.220\mbox{ GeV}^2$. As expected, the location of the left hand side pole falls between the two left hand side poles of the $K$-matrix parametrization. 
 In Fig.~\ref{fig:Ds} we show the inverse of the denominator functions  $D^{new}_{\pi\pi}$ and $D^{new}_{\pi K}$.

 \subsection{ Interpretation of the $J/\psi \rightarrow 3 \pi$ data} \label{couple}

With the left hand cut singularities of the scattering amplitudes given by a simple pole,
 ({\it cf.} Eqs.(\ref{Nnew11}),(\ref{Nnew12}))   from   Eq.(\ref{Pim}) it follows that   $Im P_\alpha(s) =0$. Thus $P_\alpha(s)$ is analytical in the entire $s$-plane and therefore given by a polynomial, 
 \begin{equation} 
 P_\alpha(s) = (s - s_{L})  C_\alpha(s). 
 \end{equation} 
The first  term is  responsible for removing the left hand cut singularities from $N^{new}_{\alpha\beta}(s)$ and making $\hat F_\alpha(s)$ in Eq.(\ref{prod}) analytical  for $s<0$. 
 The  bound $|P_\alpha(\infty)| < 1$ restricts $C_\alpha(s)$ to be at most a first order polynomial in $s$.  Thus  the final solution to Eq.(\ref{prod})  has the form 
  \begin{equation} 
  F_1(s) = N q_\pi(s) p_\pi(s)  \left[  \frac{ 1  + a_\pi s  }{D^{new}_{\pi\pi}(s)}  
  + r_{\pi K} \frac{ 1 + a_K s }{D^{new}_{\pi K}(s)}  \right].  \label{f1} 
  \end{equation} 
The first term corresponds to  $J/\psi \to (\pi\pi)_P \pi$ and the second 
  to the re-scattering contribution from   $J/\psi \to (K{\bar K})_P \pi \to (\pi\pi)_P \pi$.  

\begin{figure}[tt]
\begin{center}
\includegraphics[width=3. in,angle=00]{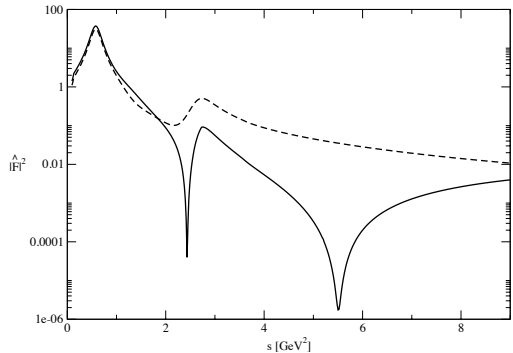} 
\caption{  The isobar form factor $|F_\pi(s)|$ with a single $\pi\pi$ channel (dashed) and with both $\pi\pi$ and $K{\bar K}$ channels (solid) using the same parameters as in Fig.~\ref{coupledalitz}.  
   \label{coupleampsq}}
\end{center}
\end{figure}


    \begin{figure}[tt]
\begin{center}
\includegraphics[width=3. in,angle=0]{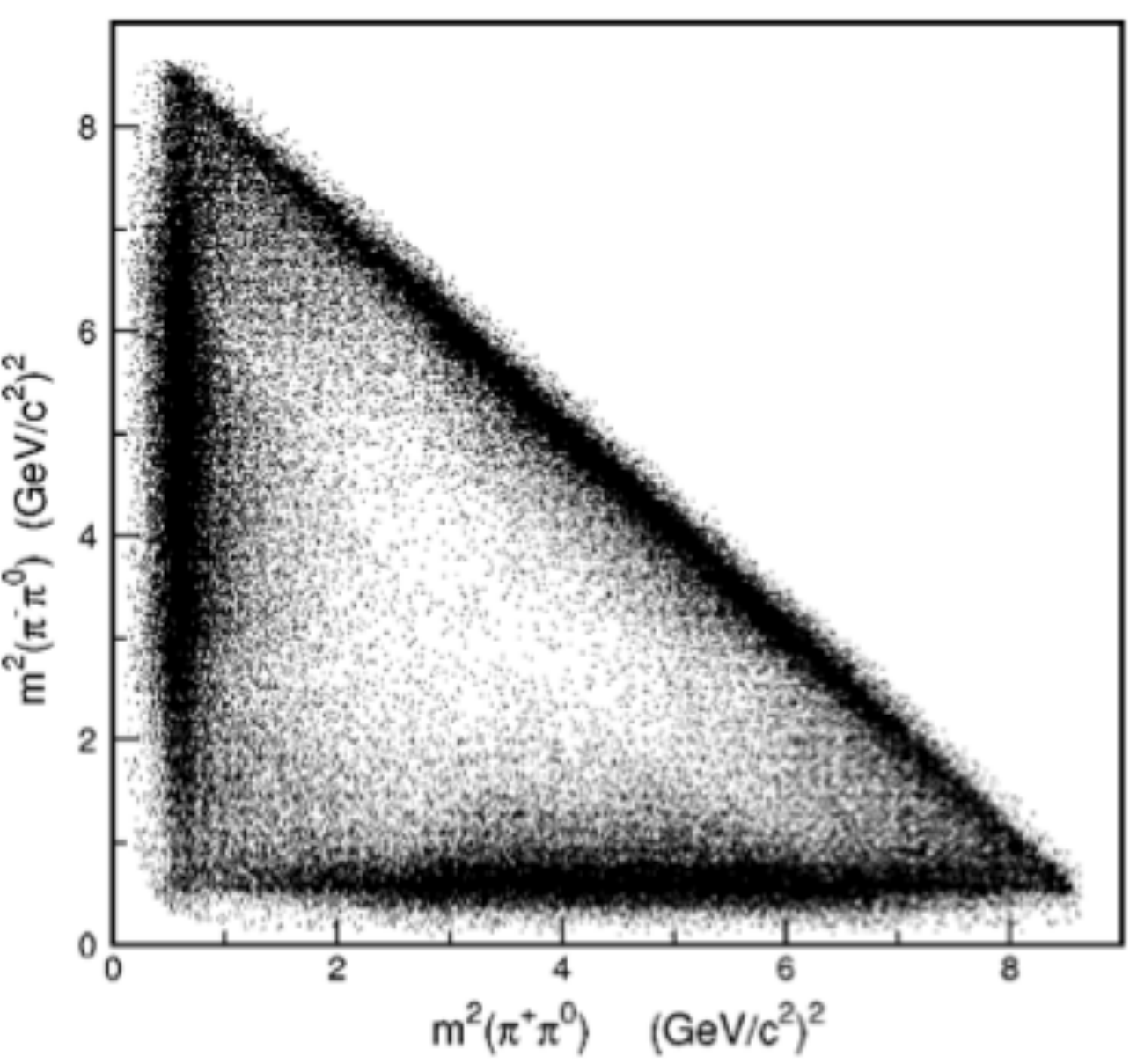}  
\caption{The $J/\psi \to \pi^+\pi^- \pi^0$  Dalitz plot  distribution 
from the BES Collaboration~\cite{Bai:2004jn}.
\label{bes}}
\end{center}
\end{figure}

  \begin{figure}[hh]
\begin{center}
\includegraphics[width=3. in,angle=90]{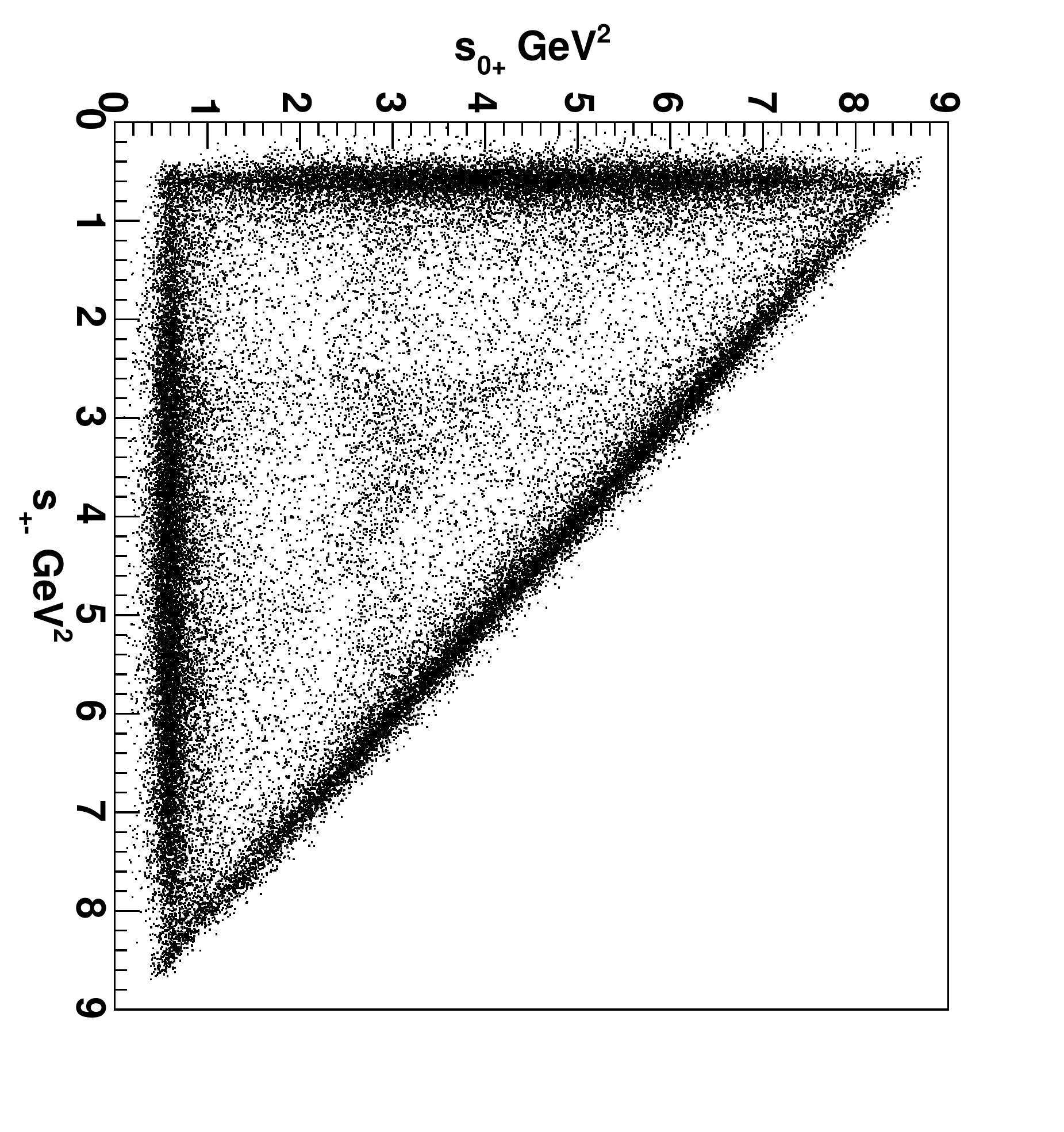}  
\caption{ Dalitz plot  distribution with the single  $\pi \pi$ channel only 
 {\it i.e.}  $\hat F_1(s) =  q_\pi p_\pi/D_{11} $ instead of Eq.(\ref{f1}). 
 \label{singledalitz}}
\end{center}
\end{figure}

\begin{figure}[hh]
\begin{center}
\includegraphics[width=3. in,angle=90]{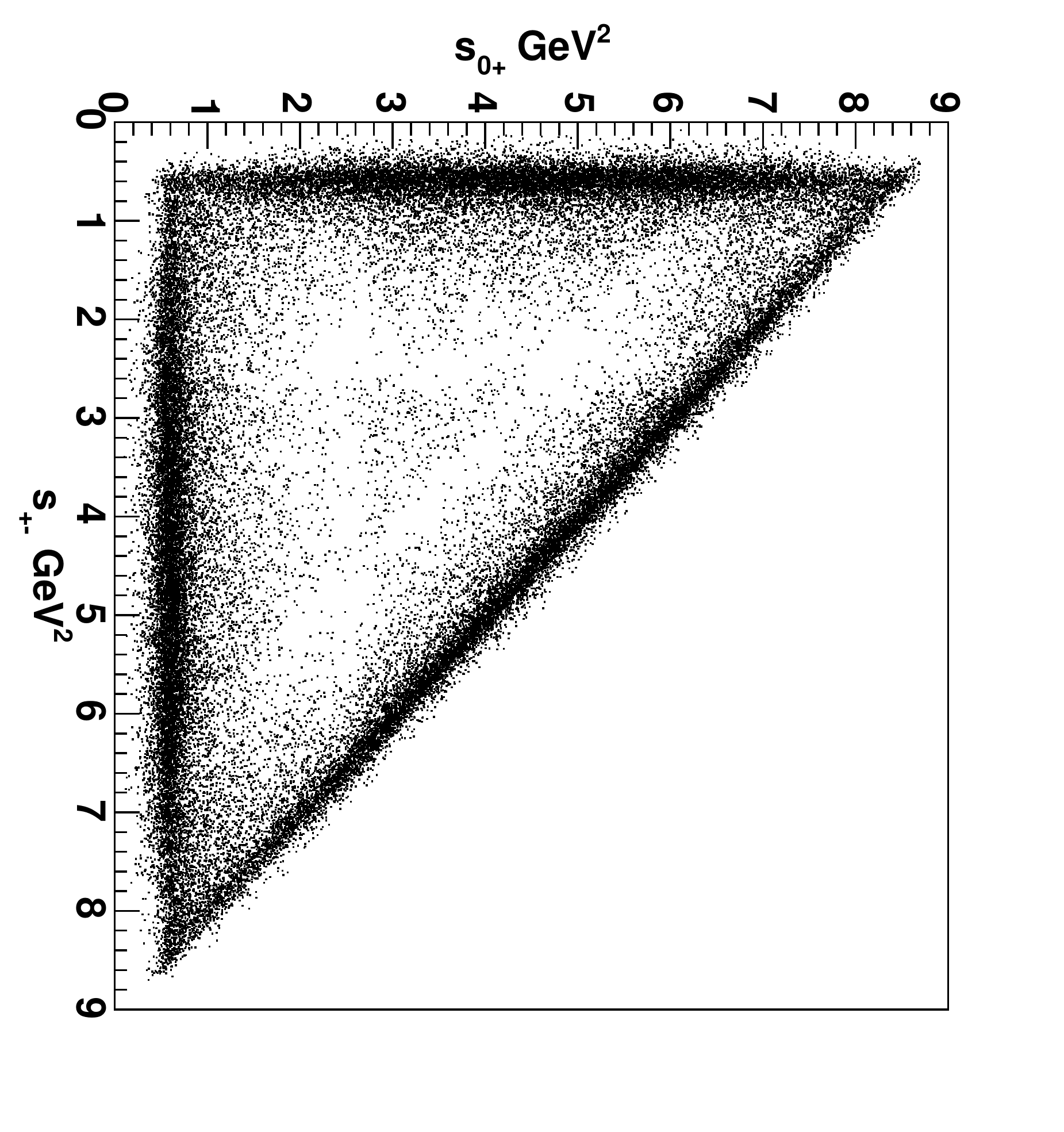}  
\caption{ Dalitz plot distribution from $\hat F_1$ in Eq.(\ref{f1}) with both the $\pi \pi$ and $K \bar{K}$ channels with   $a_\pi = -1.5 \times 10^{-1}\mbox{GeV}^{-2}$ and  $r_{\pi/K} = -1.3\times 10^{-2}$.  \label{coupledalitz}}
\end{center}
\end{figure}

        The Dalitz distribution of $3\pi$ events from $J/\psi$ decays is shown in Fig.~\ref{bes} and the striking feature is the depletion of events in the center of the plot.  
        This is to be compared with the distribution shown in Fig.~\ref{singledalitz}, which has been
          generated with $r_{\pi K}  = 0$. 
      The three bands originate from the $\rho$ meson contribution to $D^{new}_{\pi\pi}$  and the 
       large contribution from the $\rho'(1600)$ resonance 
        leads to a significant population of events  in the middle of the Dalitz  plot that is  
       not seen in the data in Fig.~\ref{bes}. Furthermore in the data there is a large contribution  near the tails of the $\rho$ bands, which are absent if only the direct $3\pi$ production is considered. 
We thus consider the full amplitude   from Eq.(\ref{f1})  and float the three parameters $a_\pi ,a_K$ and $r_{\pi K}$ 
 to obtain a distribution that best resembles the data.  We find little sensitivity to the term proportional to $a_K$ and thus set $a_K=0$. The parameter $a_\pi$ is relevant since it controls the tail of the $\rho$ resonance and so is $r_{\pi K}$ which determines the relative strength of 
   the $K\bar K$ contribution which interferes with the $\pi\pi$ amplitude in the $\rho'(1600)$ region and  reduces the contribution at the center of the Dalitz plot. 
 In Fig.~\ref{coupledalitz}  we show the event distribution using 
 $a_\pi = -1.5 \times 10^{-1} \mbox{GeV}^{-2}$ and  $r_{\pi/K} = -1.3 \times 10^{-2}$.
 
 The normalization constant $N$ is at this stage arbitrary since we are not  determining the absolute value of the branching ratio. 

Now, inspecting the  Dalitz plot  in Fig.\ref{coupledalitz} and the plot of the function 
 $|\hat F_\pi(s)|$ in Fig.\ref{coupleampsq}, it is seen that the $K\bar{K}$ channel can indeed  bring theory closer to the data by enhancing the $\pi\pi$ contribution in the energy range $1 \mbox{ GeV}  < \sqrt{s} < 1.5  \mbox{ GeV}  $ and reducing the strength of the  $\rho'(1600)$ peak.  

\section{Summary} \label{conclusion}

We have studied the effects of inelastic $\pi\pi$ scattering on the $J/\psi \rightarrow 3\pi$ Dalitz plot.  We have seen that the $K\bar{K}\rightarrow\pi\pi$ channel can significantly alter the shape of the Dalitz plot, especially at higher $\pi\pi$ masses.  This brings the observed data closer to the phenomenological expectations based on $\pi\pi$ $P$-wave scattering.  These coupled channel effects will become even more important as experimental data sets grow larger, for example at BES~III, where 1~billion $J/\psi$ decays are expected. 
 
 \section{ACKNOWLEDGMENTS  } 
 This work was supported in part by the US Department of Energy grant under 
contract DE-FG0287ER40365 and National Science Foundation PIF grant number 0653405.

\end{document}